\documentclass[trackchanges, twocolumn]{aastex701}

\usepackage{amsmath}
\usepackage{listings}
\usepackage{color}

\definecolor{dkgreen}{rgb}{0,0.6,0}
\definecolor{gray}{rgb}{0.5,0.5,0.5}
\definecolor{mauve}{rgb}{0.58,0,0.82}

\newcommand{\kms}{\hbox{km\,s$^{-1}$}}

\newcommand{\ith}{\ensuremath{^{\rm th}}}

\newcommand\totalsample{539 }

\newcommand{\amnh}{Department of Astrophysics, American Museum of Natural History, Central Park West, New York, NY, USA}
\newcommand{\planetarium}{Plan\'etarium de Montr\'eal, Espace pour la Vie, 4801 av. Pierre-de Coubertin, Montr\'eal, Qu\'ebec, Canada}
\newcommand{\irex}{Trottier Institute for Research on Exoplanets, Universit\'e de Montr\'eal, D\'epartement de Physique, C.P.~6128 Succ. Centre-ville, Montr\'eal, QC H3C~3J7, Canada}
\newcommand{\columbia}{Department of Astronomy, Columbia University, 550 West 120th Street, New York, NY 10027, USA}
\newcommand{\cuny}{Department of Physics and Astronomy, Hunter College, City University of New York, 695 Park Avenue, New York, NY 10065, USA}

\newcommand{\marianopolis}{Marianopolis College, 4873 Westmount Ave.Westmount, QC H3Y 1X9, Canada}

\newcommand{\mcgill}{McGill Institute for Aerospace Engineering (MIAE), McGill University, 817 Sherbrooke St. W.Montreal, QC H3A 0C3, Canada}
\received{June 2nd, 2026}
\revised{July 28th, 2026}
\accepted{August 18th, 2026}
\begin{document}

\title{Improving the Physical Interpretability of Gaussian Processes in Stellar Activity Modeling: A Study Case on Photometric Variability Among Stellar Clusters}

\author[orcid=0000-0001-7171-5538]{Leslie Moranta}
\affiliation{\planetarium}
\affiliation{\irex}
\affiliation{\amnh}
\email{leslie.moranta@umontreal.ca}
\email[show]{leslie.moranta@umontreal.ca}

\author[orcid=0000-0002-2592-9612]{Jonathan Gagné}
\affiliation{\planetarium}
\affiliation{\irex}
\email{gagne@astro.umontreal.ca}

\author[orcid=0000-0001-6251-0573]{Jacqueline K. Faherty}
\affiliation{\amnh}
\email{jfaherty@amnh.org}

\author[0000-0001-9482-7794]{Mark Popinchalk}
\affiliation{\amnh}
\email{popinchalkmark@gmail.com}

\author[0000-0002-2792-134X]{Jason Lee Curtis}
\affiliation{\columbia}
\email{jasoncurtis.astro@gmail.com}

\author[0000-0003-0489-1528]{Johanna M. Vos}
\affiliation{School of Physics, Trinity College Dublin, The University of Dublin, Dublin, Ireland}
\affiliation{\amnh}
\email{Johanna.Vos@tcd.ie}

\author[0009-0005-6135-6769]{Alexandrine L'Heureux}
\affiliation{\irex}
\email{alexandrine.lheureux@umontreal.ca}

\author{Andrew Ayala}
\affiliation{\amnh}
\affiliation{\cuny}
\email{andrew.ayala.2018@gmail.com}

\author{Eleanor Johnson}
\affiliation{\amnh}
\email{ejohnson1@amnh.org}

\author{Dawn McCullough}
\affiliation{\amnh}
\email{dmccullough@amnh.org}

\author{Nicole Munoz}
\affiliation{\amnh}
\email{nmunoz@amnh.org}

\author{Hannah Park}
\affiliation{\amnh}
\email{hpark@amnh.org}

\author{Xue Weng}
\email{xweng@amnh.org}

\affiliation{\amnh}
\author{Livia Poliquin}
\affiliation{\marianopolis}
\email{livpolik@gmail.com}

\author{Ilhem Lazizi}
\affiliation{\mcgill}
\email{ilhemlazizi@gmail.com}

\begin{abstract}

Gaussian Processes (GPs) are widely used to model stellar variability in photometric surveys, but a statistically successful fit does not guarantee that the inferred hyperparameters correspond to physically meaningful stellar properties. This is especially important for young, active stars, whose TESS light curves contain evolving spots, harmonics, and non-sinusoidal variability. We use stellar rotation as a case study to examine when the period hyperparameter of a quasi-periodic GP can be interpreted as a physical rotation period. We introduce a regularized GP likelihood that reweights the covariance-complexity term in the marginal likelihood, reducing the tendency of unconstrained models to converge toward preferred but misleading solutions.

We test this framework on 539 stars in IC~2602, the Tucana--Horologium Association, Pisces--Eridanus, and Group~X, with independently reported and manually vetted rotation periods. This benchmark evaluates GP hyperparameter interpretability and automated period recovery with minimal human intervention. We compare regularized and unregularized models across several regularization strengths, $\lambda$, using literature agreement, sector-level diagnostics, and consistency across TESS sectors. Relative to the standard GP likelihood, regularization improves successful rotation-period recovery by an average of 7\%. It also generally reduces the median absolute fractional deviation of periods across sectors, showing that the improvement is not limited to catastrophic failures but also mitigates smaller systematic errors.

Regularization is particularly beneficial for non-sinusoidal or evolving modulation, where unconstrained GPs may recover harmonics or spurious timescales. These results show that likelihood regularization and cross-sector consistency are practical diagnostics for assessing when GP-based rotation periods are robust.

%To further validate the method, we apply the same framework to simulated light curves generated with the \texttt{butterpy} package and obtain comparable improvements in recovery performance. 

\end{abstract}

\section*{Introduction}\label{sec:intro}
Historically, young stars in the Solar neighborhood were detected due to strong X-ray radiation \citep{Carkner1997}. However, decades of research revealed that even the seemingly isolated young stars reside in associations with hundreds to thousands of members. These associations form during the collapse of a single molecular cloud \citep{Lada2003}. Consequently, stars in such associations share similarities useful for both identifying members and studying early stellar evolution. Most members of these young associations and clusters are comoving (of similar Galactic space velocities within $\approx 0.5$\,\kms) and coeval (of the same age within a few Myr).

Precise astrometry from the Gaia missions \citep{Gaia2016,Gaia2018,Gaia2021}, combined with clustering algorithms such as \texttt{HDBSCAN}  \citep{Campello2013}, allowed the discovery of many new comoving stars in the solar neighborhood \citep{Kounkel2019, Moranta2022}. Young clusters disperse over time, eventually becoming indistinguishable from older field stars. Consequently, only associations younger than $\approx$ 1 Gyr, or denser, gravitationally bound open clusters, are typically detectable \citep{Dinnbier2020}.

Determining the age of individual stars is challenging because age-dating methods are usually indirect, they rely on model assumptions, or they provide strong constraints only with specific types of stars \citep{Soderblom2010}. However, studying coeval stellar populations, such as comoving stars, provides a statistically robust age estimate, in large part because several age indicators can be combined into a more precise age constraint. Empirical modeling techniques, such as isochrone fitting \citep{Laughlin1997} and gyrochronology \citep{Barnes2003}, are some of the most-used methods for precise cluster age determination because they rely on readily available data.

% In contrast, the detection of H$\alpha$ and \ion{Ca}{2} lines \citep{Duncan1991,Lyra2005}, and the lithium depletion of individual stars of spectral types FGK \citep{Rebolo1991}, serve primarily as youth indicators, rather than providing precise age measurements. Some of the most accurate age-dating methods rely on determining the boundary at which the lithium reappears in the atmospheres of cooler stars (the Lithium Depletion Boundary or LDB), in part because the location of the LDB is not strongly model-dependent. However, measuring the lithium equivalent width in cool, low-mass members requires a more significant observing effort and the identification of the low-mass members can be a challenge in more-distant associations. For young stellar associations (ages $\sim$20–200 Myr), the LDB occurs at progressively cooler temperatures with increasing age. In this regime, the method yields the most accurate ages for members with masses $\gtrsim 0.3,M_{\odot}$ \citep{Burke2004,Gagne2024}.

Isochrones determine cluster ages by comparing observed star brightness and color to theoretical evolutionary tracks on a color--magnitude diagram (CMD). It remains one of the most straightforward approaches to determine the age of a cluster. However, interstellar extinction and reddening, unresolved binaries, and non-solar metallicities can affect the position of a star in the CMD and bias the recovered isochrone age. Furthermore, the isochrone ages of low-mass stars are often biased toward younger ages because of an incomplete treatment of magnetic fields in most evolutionary models \citep[e.g., see][]{Malo2014}.

Building on the work of \cite{Skumanich1972} showing stellar rotation as a youth indicator, \cite{Barnes2003} demonstrated that the rotation period of coeval stars correlates with color or effective temperature, similar to isochrone fitting \citep[e.g.,][]{Bouma2023,Curtis2019,Curtis2020,Popinchalk2023}. For F-, G-, and K-type stars aged 100 Myr to a few Gyr, gyrochronology becomes a particularly precise tool for age determination \citep{Bouma2023}. This is because these stars exhibit a progressively tighter sequence in the distribution of color and rotation period, a phenomenon driven by the gradual loss of angular momentum through stellar winds and coronal mass ejections interacting with the stellar magnetic field. The efficiency of this magnetic braking, which is generated within the convective envelope, allows for a more reliable age--rotation relationship for stars possessing such outer convective regions. The foundational empirical relation, first established by \cite{Skumanich1972} based on G-type stars observations, is applicable to F- and K-type stars as well due to their similar internal structure featuring an outer convective envelope and an inner radiative core. Recent efforts to identify nearby young stellar clusters, such as \cite{Moranta2022}, have leveraged Gaia's full astrometric solution, including radial velocity measurements. These surveys have uncovered new associations containing substantial populations of F-, G-, and K-type stars, making them well suited for the application of gyrochronology.

High stellar activity creates strong magnetic fields within a star's convection zone. These fields impede heat flow to the photosphere, resulting in cooler, darker star spots. As a star rotates, these spots cause variations in its brightness, which are detectable in photometric data \citep{Barnes2003}. The photometric measurements provided by the Transiting Exoplanet Survey Satellite \citep[TESS;][]{Ricker2014} furthered our ability to measure stellar rotation across large stellar populations. Though designed to find exoplanets, TESS continuously observes billions of nearby stars, providing a 27-day uninterrupted observation with a high cadence photometric time series.

The abundance of photometric data allows for precise gyrochronological age determination of many nearby moving groups through stellar rotation rates. However, extracting these rotation rates with unsupervised numerical methods remains challenging. As modern surveys continue to expand in size, there is an increasing need for automated pipelines capable of processing large photometric datasets. To date, most large-scale rotation studies have relied on periodogram-based approaches \citep[e.g.][]{Claytor2022, Colman2024, Boyle2026}, while fully automated Gaussian Process frameworks remain relatively unexplored. Gaussian Processes \citep[hereafter GPs;][]{Rasmussen2006}, flexible non-parametric models, are commonly used to infer periodic signals from photometric time series, like stellar rotation. Their ability to model complex signals is advantageous, but careful implementation is necessary to ensure physically meaningful results. Improperly applying GPs can yield unrealistic parameter values and hyperparameters that are difficult to interpret. To address this, we propose introducing an entropy regularization term to the GP likelihood equation that favors physically realistic results when sampling the GPs hyperparameters to model photometric data. 

In this work, we address the overfitting tendencies of Gaussian Processes in high-precision TESS photometry by introducing an explicit regularization term to the marginal likelihood, rebalancing the trade-off between model complexity and data fidelity. We describe this regularized framework and the \texttt{RotationTerm} covariance structure in Section~\ref{sec:method}. Section~\ref{sec:data} details our validation using a benchmark sample of \totalsample stars across four young populations—IC~2602, the Tucana--Horologium Association, Pisces--Eridanus, and Group~X. We present the improvement in rotation period recovery and characterize the performance of various regularization strengths in Section~\ref{sec:results}, followed by a discussion of more challenging cases for automated rotation detection using GPs in Section~\ref{sec:discussion}.
%—alongside synthetic light curves generated with \texttt{butterpy}

%%%
%%%
% METHOD SECTION 
%%%
%%%
\section{Method} \label{sec:method}

The abundance of data from the TESS all-sky survey has motivated community efforts to develop automated methods to analyze photometric data, particularly to detect stellar rotation and star spot-induced brightness variations. While numerous pipelines are currently available to detect periodic signals in light curves, no single method reliably identifies the true rotation period across a wide range of light curves without human supervision.

\subsection{Main rotation detection methods}

The Lomb--Scargle (LS) periodogram, a widely used method, estimates the Fourier transform of unevenly sampled data. While fast and efficient for determining signal periodicity in astrophysics (e.g., radial velocity, photometric variations), LS has its limitations. Specifically, LS is inefficient at recovering accurate rotation periods when multiple star spots are present, often detecting harmonics instead of the true stellar rotation period. Furthermore, although studies have attempted to estimate rotation measurement errors (e.g., equations detailed in \citet{Lamm2004,Messina2010}, or time series bootstrapping \citep{Boyle2023}), periodogram-based methods do not provide statistically robust uncertainty estimates.

Similarly, the autocorrelation function, initially used by \cite{McQuillan2013}, shows similar limitations to the LS periodogram, despite efforts to quantify uncertainties (e.g. \cite{Holcomb2022}). It measures the correlation of a signal with itself at varying time shifts, also called lags, resulting in a peak signal at the stellar rotation period.

None of the aforementioned algorithms attempt to model the data: rather, they directly extract the periodicity from the data. While periodicity is an important physical property of a star, it is also not the only information that can be extracted from a star's light curve. Other properties, such as the amplitude of the signal and the timescale of starspot evolution, are available in most light curves. Gaussian Processes \citep{Rasmussen2006} can also model these properties, making them a valuable tool for photometric data analysis.

\subsection{Gaussian Processes}

This section provides a brief overview of GPs as a numerical method, including their assumptions and how they are currently used to model and interpret astrophysical signals in spectroscopic and photometric time series. We will also discuss the limitations of GPs, potential pitfalls in the physical interpretation of their hyperparameters, and how our proposed method addresses these issues. 

GPs aim to model the underlying functional distribution of the data by assuming that they are drawn from a Gaussian distribution, as described by Equation \ref{eq:gp}.

\begin{equation} \label{eq:gp}
\scriptscriptstyle
\left[ \begin{array}{c} y_1  \\ \vdots \\ y_n \end{array} \right] \sim N \left( \underbrace{ \mu(x) }_{mean}, \underbrace{\left[ \begin{array}{ccc} \kappa(x_1, x_1) + \sigma_{y_1}^2 & \cdots & \kappa(x_1, x_n) \\  \vdots & \ddots & \vdots \\ \kappa(x_n, x_1)  & \cdots & \kappa(x_n, x_n)+ \sigma_{y_n}^2 \end{array} \right]}_{\text{covariance $\Sigma$}}  \right)
\end{equation}

\noindent where $\mu(x)$ denotes the mean function of the Gaussian distribution, and $\kappa(x_i, x_j)$ is the covariance function that quantifies the correlation between any two data points. The full covariance matrix, including the contribution from measurement uncertainties, is denoted by $\boldsymbol{\Sigma}$. This matrix is constructed by evaluating the covariance function at the observation times ${x_i}$. The functional form of the covariance function is referred to as the kernel, and its free parameters are known as hyperparameters.

Kernel functions can take a wide variety of forms, depending on the characteristics of the data being modeled. For example, quasi-periodic kernels are well suited for describing damped, oscillatory signals, while Matérn $3/2$ kernels are commonly used to model stochastic variability such as correlated (red) noise. More complex covariance structures can be constructed by combining multiple kernels, enabling flexible modeling of astrophysical time series data.

Young stars frequently exhibit multiple, simultaneously evolving surface features, such as starspots, which can produce harmonic or multi-periodic photometric signals. 

A key advantage of GPs is their ability to predict unseen data. Based on the Gaussian assumption, GPs predict the mean $\mu_{x_* \mid x_i}$ and covariance $\kappa_{x_* \mid x_i}$ of unseen values of $x_{*}$ based on our known values of $x_i$. The Multivariate Gaussian Theorem allows us to analytically calculate the predicted values $x_{*}$ conditioned on $x_{i}$, yielding the following equations:
\begin{equation}
\mu_{x_* \mid x_i} = \mu_{x_*} + \kappa_{x_*, x_i} \kappa_{x_i, x_j}^{-1} (x_i - \mu_{x_i})  
\end{equation}

\begin{equation}
\kappa_{x_* \mid x_i} = \kappa_{x_*, x_*} - \kappa_{x_*, x_i} \kappa_{x_i, x_j}^{-1} \kappa_{x_i, x_*}
\end{equation}

In astrophysical time-series data analysis, Gaussian Processes are typically not employed to interpolate or forecast missing data, but rather to remove noise and to extract physically meaningful information from the inferred covariance structure. To this end, GP hyperparameters are commonly estimated using sampling-based inference techniques, such as Markov Chain Monte Carlo, Hamiltonian Monte Carlo, or nested sampling. The resulting posterior distributions of the hyperparameters are then interpreted in terms of underlying physical processes.

For example, GPs have become increasingly popular in recent years to model stellar activity in both photometric and spectroscopic data \citep{Rajpaul2015,Aigrain2016, Angus2018}. When using a suitable kernel, it becomes possible to translate a GP hyperparameter into a stellar property. For example, the hyperparameter $P$ may correspond to the rotation period of a star, the length scale (a transformation of $Q_0$) can correspond to the lifetime of a stellar spot, and $\sigma$ can correspond to the amplitude of the photometric variability.

The posterior likelihood $\mathcal{L}$ associated with a set of hyperparameters, given an observed data set $y$ can be written as:

\begin{equation}\label{eq:like}
  \log \hat{\mathcal{L}}  =  \underbrace{- \dfrac{1}{2N} \log{\mid \Sigma \mid} }_{\text{Complexity Penalty}} - \underbrace{ \dfrac{1}{2N} \textbf{y}^T \Sigma^{-1} \textbf{y} }_{\text{Data-fit}}  - const.
\end{equation} 

The log marginal likelihood in Equation~\ref{eq:like} can be decomposed into two distinct contributions: a data-fitting term and a complexity penalty term, the latter acting as an automatic regularization of the model complexity \citep{Rasmussen2006}. While the data-fitting term rewards models that closely reproduce the observed flux, the complexity term penalizes overly flexible covariance structures by favoring simpler models.

\begin{figure*}[ht!]
\plotone{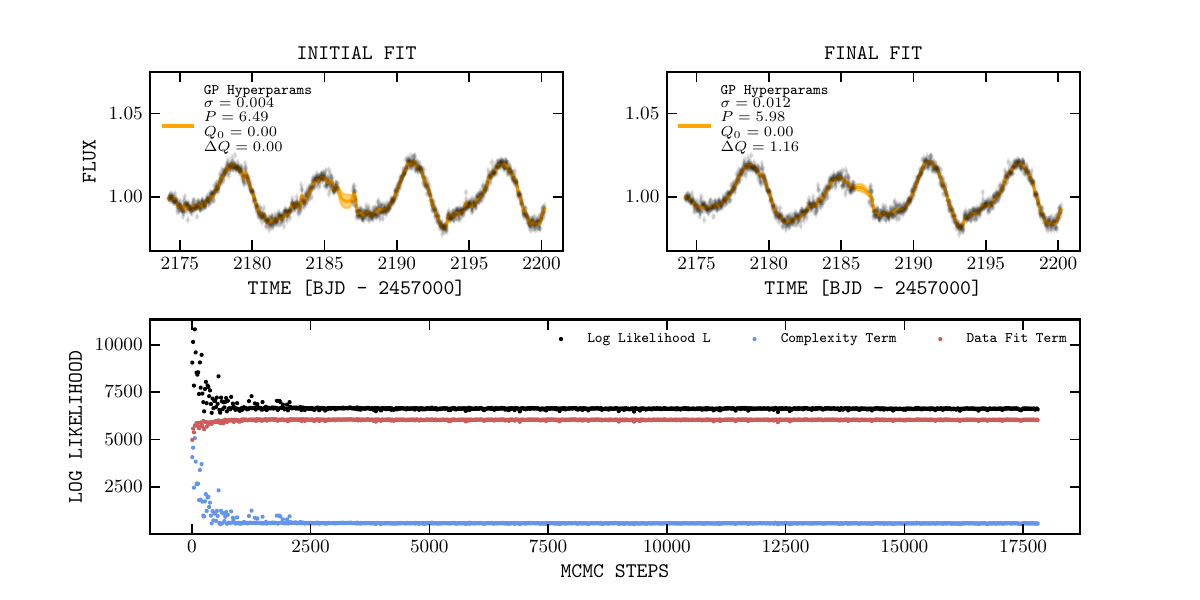} 
\caption{
TESS Sector~32 light curve of Gaia DR3 4784114803446931200, a member of the Tucana--Horologium Association, modeled using a Gaussian Process.
\textit{Top left:} GP model evaluated using the initial set of hyperparameters at the start of the MCMC sampling. Despite the lack of optimization, the model already reproduces the observed variability.
\textit{Top right:} GP model corresponding to the maximum-posterior hyperparameters obtained after MCMC convergence.
\textit{Bottom:} Evolution of the binned log marginal likelihood terms during a 50-walker MCMC run. Blue points show the complexity penalty term of Equation~\ref{eq:like}, red points show the data-fitting term, and black points show the total log likelihood. The data-fitting term remains nearly constant throughout the sampling, while the overall likelihood variations are dominated by the complexity penalty, indicating that hyperparameter inference is primarily driven by model complexity rather than data fit.}
\label{fig:like_eval}
\end{figure*}

In the Gaussian Process framework, a distinction is made between the latent function and the hyperparameters that define its covariance structure. The GP models the data as a realization of an underlying latent function, whose behavior is governed by a chosen covariance kernel. The hyperparameters of this kernel control properties such as the amplitude, characteristic timescales, and periodicity of the signal. In this work, we adopt a quasi-periodic kernel (Equations 2 and 3), in which the rotation period is encoded in one of the hyperparameters. Alternative kernel choices are possible, but this form is commonly used to model stellar variability. For a given set of hyperparameters, many realizations of the latent function can provide similarly good fits to the data, as illustrated in the top panels of Figure~\ref{fig:like_eval}. As a result, the data-fitting term alone provides only weak constraints on the hyperparameters.

When training the GP on the full TESS light curve, particularly in the regime of small photometric uncertainties, we find that the dominant contribution to the log likelihood arises from the complexity penalty term (Figure~\ref{fig:like_eval}, bottom panel, see \textit{Complexity penalty} term in Equation~\ref{eq:like}). In this limit, hyperparameter optimization effectively reduces to identifying the simplest covariance structure that passes through each data point within the quoted uncertainties. While this behavior is mathematically well defined, it does not guarantee the physical interpretation of the inferred hyperparameters.

Photometric time series data encodes a superposition of multiple astrophysical processes, including stellar rotation, starspots, flares, planetary transits, contamination from background eclipsing binaries, etc. Consequently, the simplest GP model that explains the data need not correspond to any single physical mechanism. The inferred kernel timescales and amplitudes should therefore not be interpreted as direct measurements of stellar rotation or activity without additional physical modeling or explicit separation of these signals. In this sense, GP hyperparameters reflect statistical properties of the data rather than uniquely identifiable physical quantities. 

\subsection{Regularization}

By construction, regularization enforces a trade-off between model flexibility and fidelity to the data, discouraging overly complex solutions that explain noise rather than underlying structure. In the context of GPs, this trade-off is implicitly encoded in the marginal likelihood through the complexity penalty term. However, as shown in Figure~\ref{fig:like_eval}, when modeling high-precision photometric time series data such as TESS data, the likelihood can be driven by the complexity term without ever affecting or allowing flexibility in the data fitting term, forcing the model to go through each point.

In many machine learning applications, regularization strength is explicitly controlled through a weighting parameter that is tuned using validation techniques such as cross-validation or leave-one-out tests \citep{Stone1974}. By contrast, astronomical applications of GPs have largely relied on the standard likelihood optimization, without systematically exploring how regularization may mitigate overfitting or improve physical inference. As a result, GP hyperparameters often converge to values that favor the simplest model consistent with all data points, rather than those that best capture the dominant stellar signal.

Motivated by these considerations, we modify the GP log marginal likelihood by introducing an explicit weighting factor that re-balances the contribution of the data-fitting term relative to the complexity penalty. The resulting likelihood function is:
\begin{equation}\label{eq:like_modified}
\log \hat{\mathcal{L}}_{new} =
\underbrace{ -\dfrac{\lambda}{2N} \log \left| \boldsymbol{\Sigma} \right| }_{\text{Complexity penalty}}
-
\underbrace{ \dfrac{1}{2N} \mathbf{y}^T \boldsymbol{\Sigma}^{-1} \mathbf{y} }_{\text{Data fit}} ,
\end{equation}

\noindent where $\lambda \geq 1$ controls the relative importance of the data-fitting term. This parameter is not fixed \emph{a priori} and must be tuned to achieve an optimal balance between robustness to outliers and sensitivity to astrophysically meaningful variability. In Appendix~\ref{app:entropy}, we discuss the connection between this formulation and entropy-based regularization, and provide additional intuition for the interpretation of $\lambda$ in the context of GP modeling.

\subsection{Practical Implementation}
In young stars, photometric time series data often exhibit a rich combination of astrophysical variability, including flares, evolving starspots, and instrumental systematics. Nevertheless, the dominant signal is frequently the quasi-periodic modulation induced by surface inhomogeneities carried across the stellar disk by rotation. Our goal is to demonstrate that, by combining a physically motivated covariance structure with explicit regularization, GP modeling can yield hyperparameters that more robustly trace the stellar rotation rate.

To accurately characterize the stellar rotation signal, we adopt the \texttt{RotationTerm} covariance structure implemented in the \texttt{celerite2} package \citep{celerite2}, which is defined as the sum of two stochastically driven, damped harmonic oscillators. The primary component of this kernel is given by:

\begin{equation*}
     Q_1 = 1/2 + Q_0 +\delta Q
\end{equation*}

\begin{equation*}
     \omega_1 = \dfrac{4 \pi Q_1}{P \sqrt{4Q_1^2-1}}
\end{equation*}

\begin{equation}
     S_1 = \dfrac{\sigma^2}{\sqrt{(1+f)\omega_1 Q_1}}
\end{equation}

and the secondary term is described by:

\begin{equation*}
     Q_2 = 1/2 + Q_0 
\end{equation*}

\begin{equation*}
     \omega_2 = \dfrac{8 \pi Q_2}{P \sqrt{4Q_2^2-1}}
\end{equation*}

\begin{equation}
     S_2 = \dfrac{f \sigma^2}{\sqrt{(1+f)\omega_2 Q_2}}
\end{equation}

\noindent where $\sigma$ (amplitude), $P$ (period), $\delta Q$ (difference of quality factor between the rotation and its harmonic), $Q_0$ (quality factor of the secondary signal), and $f$ (fraction of amplitude of the secondary signal) are the hyperparameters of the model.

We apply our modified likelihood formulation to this kernel in order to control the balance between data fidelity and model complexity, with the aim of improving the physical interpretability of the inferred hyperparameters. More specifically, we assess whether the inferred rotation rates are consistent with independent measurements from the studies in \ref{sec:data} using TESS data.

For all experiments, the prior distributions assigned to each GP hyperparameter are listed in Table~\ref{tab:priors}. These priors are held fixed across different regularization strengths to isolate the impact of the likelihood weighting on the inferred rotation parameters.

\tablewidth{1\textwidth} 
\begin{deluxetable}{ccc}[!ht]\label{tab:priors}
\tablecolumns{2}
\tablecaption{Priors to the GP \texttt{RotationTerm} model from \texttt{celerite2}.}
\tablehead{\colhead{Parameter} &  \colhead{Prior}} 
\startdata
$\sigma$ &  $\mathcal{L}\mathcal{U}(10^{-5}, 1)$\\
$P$ & $\mathcal{L}\mathcal{U}(0.09, 15)$ \\
$Q_0$ & $\mathcal{L}\mathcal{U}(10^{-5}, 10^5)$ \\
$dQ$ &  $\mathcal{L}\mathcal{U}(5, 100)$\\
$f$ & $\mathcal{L}\mathcal{U}(10^{-5}, 0.7)$ \\
$wn$ & $\mathcal{L}\mathcal{U}(10^{-5}, 0.1)$\\
\enddata
\end{deluxetable}

As described in Appendix~\ref{sec:packages_m}, we implement the modified likelihood by introducing a weighting factor $\lambda$, which scales the complexity term. We explore values of $\lambda = 1, 2, 4, 8,$ and $16$, where $\lambda = 1$ corresponds to the standard GP marginal likelihood. For each value of $\lambda$, we perform full posterior sampling of the GP hyperparameters.

In Section~\ref{sec:data}, we describe the dataset used to evaluate the performance of this pipeline and compare the rotation rates inferred from the GP hyperparameters against values reported in the literature.

%%%
%%%
% DATA SECTION 
%%%
%%%
\section{Data}\label{sec:data}
We validated our rotation rate detection method by comparing its results to published rotation rates for stars in nearby clusters. We also removed stars whose TESS light curves are potentially contaminated by other nearby young stars (described in Section \ref{sec:method}). We selected benchmark clusters across a wide range of ages that benefit from good TESS coverage.

We only used rotation rates determined through manual inspection to ensure high-quality comparisons. These criteria yielded the following benchmark clusters: The Tucana--Horologium Association  \citep{Zuckerman2000} and IC\,2602 \citep{Markarian1953,Hoogerwerf1999,vanLeeuwen1999,Nisak2022} open cluster ($\sim 40$ Myr), the Pisces--Eridanus stellar stream \citep{Meingast2019} ($\approx 120$ Myr), and Group-X \citep{Oh2017} ($\sim 300$ Myr). Not only does this sample include \totalsample stars with manually determined rotation rates from TESS data, but they span a critical age range (40–300 Myr) where stars are transitioning from rapid to slow rotation.

The following subsections detail the rotation measurements for each member of these clusters. 

\subsection{The Tucana--Horologium Association and IC 2602 Open Cluster}

The Tucana--Horologium Association  \citep{Zuckerman2000}, located within 100 pc and with an age of $\sim$40 Myr, is currently thought to have formed directly as a loose population (e.g., \cite{Jerabkova2019}) in the same star formation event and molecular cloud complex as the IC~2602 open cluster \citep{Gagne2021,Popinchalk2024}. \cite{Gagne2021} named this the IC~2602 system, which includes the IC~2602 core, its surrounding halo (CIC~2602), and a tidal tail comprised of Tucana--Horologium Association members. This hypothesis was derived from kinematic features and indications of coevality. \cite{Popinchalk2024} further confirmed their coevality by showing an overlap in their gyrochronological ages.

\cite{Popinchalk2023} performed a photometric analysis of light curves using the methods outlined in \cite{Curtis2019} and \cite{Popinchalk2023,Popinchalk2024}. They extracted Full Frame Image (FFI) cutouts of 40 $\times$ 40 pixels \citep[\texttt{TESScut};][]{Brasseur2019}. Two detrending methods are used in \cite{Popinchalk2023} to reduce systematic noise: the Causal Pixel Model (CPM), implemented in \texttt{unpopular} \cite{Hattori2021}, which uses pixels outside the target aperture within the same charge-coupled device (CCD), and Simple Aperture Photometry (SAP), which subtracts background signal from the target pixel. As discussed in \cite{Popinchalk2023}, CPM is better suited for faint objects ($G_{Mag} \geq 10$), while SAP is more appropriate for bright objects ($G_{Mag} \leq 10$). In \cite{Popinchalk2023}, every TESS sector available (1 - 65 at the time of their publication) is inspected for each target by using the Lomb--Scargle periodogram \citep{VanderPlas2018}. The sectors that provided the most comprehensive representation of the photometric variability of the stars are then utilized to derive the final classification of the object. They used four possible classifications:
\begin{itemize}
\item {Publish:} The light curve shows a clear and powerful periodic signal. In this case, they adopt the signal as the rotation rate of a star.
\item {Good:} The light curve shows a periodic signal, but not as strong as those classified in Publish. In this case, they adopt the signal as the rotation rate of a star.
\item {Flat:} The light curve of the object exhibits no sign of photometric variability. In this case, they do not provide a rotation rate  to the object.
\item {Garbage:} Signal is dominated by instrument systematics or detrending issues. In this case, no rotation rate is given to the object.
\end{itemize}
The Tucana--Horologium Association sample initially comprised 244 objects flagged as either Good or Publish by \cite{Popinchalk2023}. After we exclude 16 contaminated targets (our contamination criteria is further described in Section~\ref{sec:results}) and two stars for which we present a revised rotation period using the rotation rate recovery method presented in this paper, 226 stars remained. Similarly, the IC~2602 sample of \cite{Popinchalk2024} is reduced from 106 to 102 stars following the removal of four contaminated objects. 
\subsection{Pisces--Eridanus}
The Pisces--Eridanus stellar stream (hereafter PERI), located between 70 and 160 pc from the Sun, is a nearby cluster first discovered by \citet{Meingast2019} using Gaia~DR2  data \citep{Gaia2018}. They initially estimated its age to be $\sim 1$ Gyr, a surprising result because most stellar stream members have dispersed by that age, and only a handful of such associations have been discovered (e.g., NGC 6811, Ruprecht 147). However, \citet{Curtis2019} studied the group with TESS and revised the age to $\sim 120$ Myr using gyrochronology applied to FGK members and isochrone fitting of high-mass candidate members. This revision demonstrated that the gyrochronological age of PERI is consistent with the Pleiades.
For Pisces--Eridanus, the initial sample includes 100 stars, of which 98 are retained in the final analysis after excluding two targets for which we derived updated rotation periods with our GP analysis. For consistency purposes, our analysis used only TESS sectors 1--5 to parallel the initial study by \cite{Curtis2019}.
\subsection{Group-X}
Group~X, located between 90 and 110 pc from the Sun, is a nearby moving group first identified by \cite{Oh2017} using Gaia DR1 \citep{Gaia2016} data, with an initial sample of 27 candidate members. Subsequent studies, including \cite{Faherty2018} and \cite{Tang2019}, expanded the membership list to 218 candidate members, and confirmed that Group~X is distinct from the nearby Coma Berenices cluster \citep{Melotte1915,Trumpler1938}. Group~X is undergoing rapid dispersal, as indicated by its kinematic morphology. \cite{Tang2019} estimated an age of $\approx$\,400 Myr for Group~X based on a comparison of its Gaia~DR2 color--magnitude diagram with the PARSEC isochrones of \citep{Bressan2012}. More recent studies combining stellar rotation and lithium measurements instead indicate a younger age of $\sim300$\,Myr \citep{Newton2022,Messina2022}.
In this work, we adopt the rotation periods measured by \citet{Newton2022}, who performed a detailed analysis of Group~X using both TESS and ZTF photometry. Their study combined the membership catalogs of \citet{Tang2019} and \citet{Furnkranz2019} with additional candidate members identified through the \texttt{FindFriends} algorithm \citep{Tofflemire2021}, resulting in an expanded sample of candidate Group~X stars.
A key advantage of the \cite{Newton2022} analysis is their careful treatment of TESS systematics. They demonstrated that the pre-search data conditioning simple aperture photometry (PDCSAP) light curves can artificially suppress long-period rotational variability and bias some stars toward half-period solutions, particularly for stars with rotation periods between $\sim6$ and 12 days. To mitigate this issue, they instead adopted SAP light curves and custom full-frame image reductions, which better preserved the intrinsic long-timescale stellar variability relevant for gyrochronology.
Rotation periods were measured using Lomb--Scargle periodograms applied to both TESS and the Zwicky Transient Facility (ZTF) photometry, with candidate detections vetted through visual inspection of the light curves, phased curves, and autocorrelation functions. Notably, the authors assessed the reliability of the detected rotation periods by visually vetting each candidate signal and required that:
\begin{enumerate}
    \item The peak be clearly identifiable in the Lomb--Scargle periodogram;
    \item The phase-folded light curve display a coherent quasi-periodic repeating modulation consistent with stellar rotation.
\end{enumerate}
Out of consistency with the other surveys, we only considered stars with TESS rotation detection, for which \cite{Newton2022} ultimately reported reliable rotation periods for 112 candidate Group~X members using the 27 TESS sectors available at the time.
% The gyrochronological study of Group X by \cite{Messina2022} refined its age to $\approx 300$\,Myr by analysing the TESS light curves detrended by the \textit{PSF-based Approach to TESS High quality data Of Stellar clusters} (PATHOS) pipeline \citep{Nardiello2019}, and assigned confidence grades to their measured rotation periods based on how consistently the values were found using three different methods: (1) Generalized Lomb-Scargle (GLS) \citep{Zechmeister2009}, (2) time series analysis based on the \texttt{CLEAN} library \citep{Roberts1987}] and (3) the autocorrelation Function (ACF) method \citep{McQuillan2013}. They assigned the following grades:
% \begin{itemize}
%     \item Grade A: All three methods produced similar rotation periods within the uncertainties.
%     \item Grade B: Two of the three methods produced a consistent period measurement.
%     \item Unreliable (not considered): All three methods produced different measurements.
% \end{itemize}
% Additional criteria were also used to assess the detection of periodic signals, such as the consistency of rotation period across multiple TESS sectors, probing only periods with a false alarm probability $< 0.1 \%$.
\begin{figure}[ht]
\includegraphics[width=8.5cm]{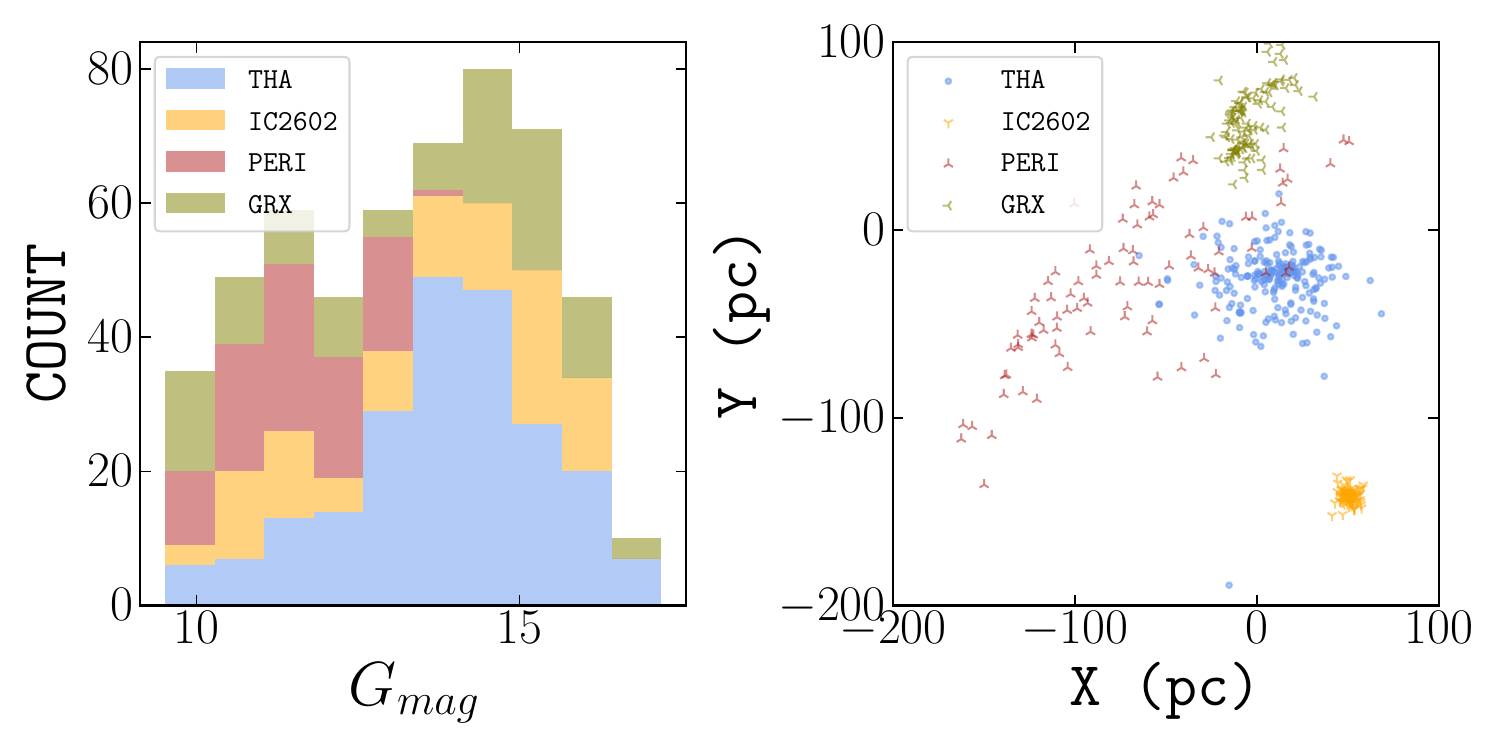}
\caption{\textit{Left:} Sample distribution for each cluster in $G_{mag}$. \textit{Right:} Sample distribution in Galactic Cartesian $X$ and $Y$ positions.}
\end{figure}
\subsection{Photometric Data}\label{subsec:phot}

\begin{figure}[ht] 
\includegraphics[width=8.5cm]{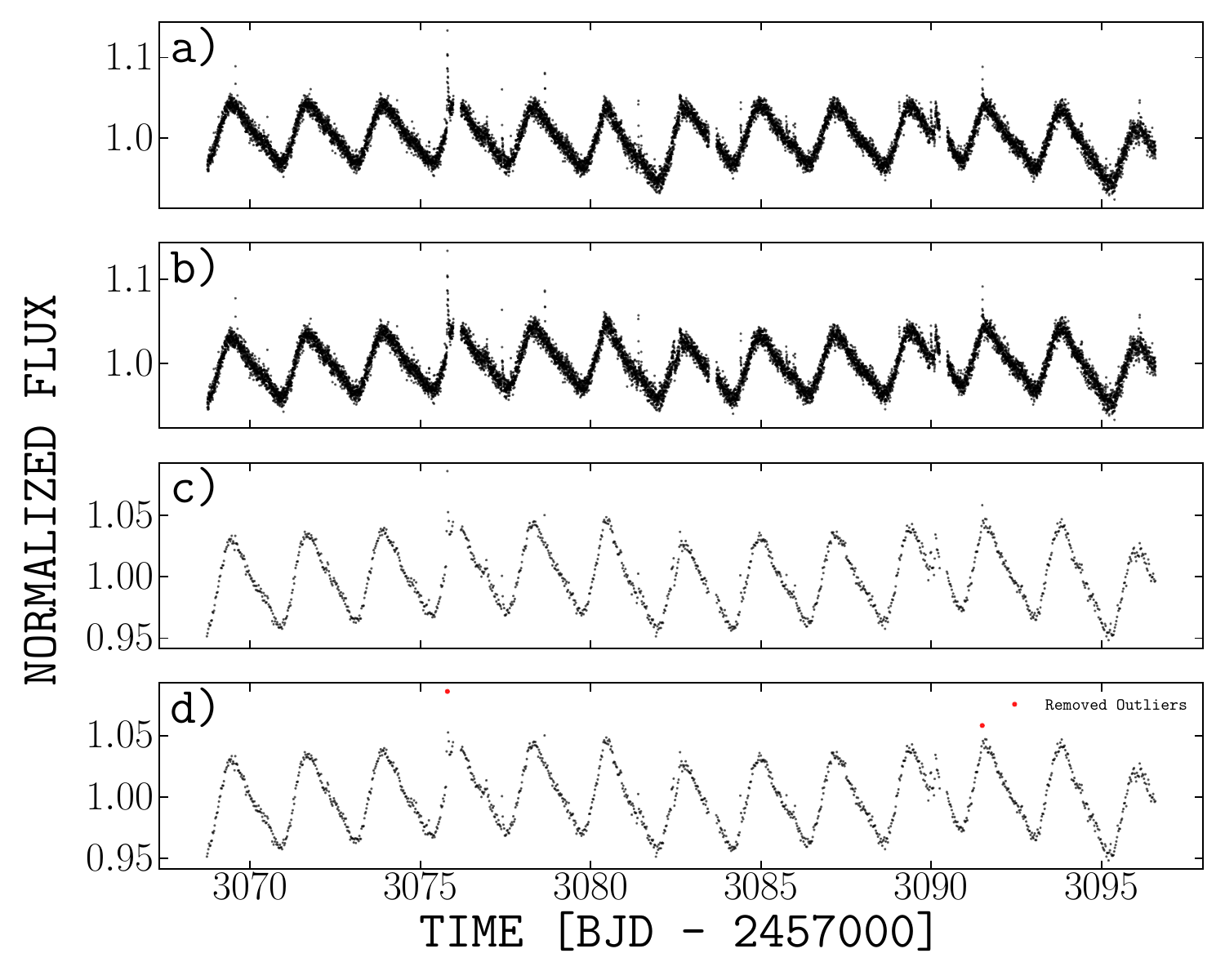}
\caption{
\emph{TESS} light curve of the Tucana--Horologium Association member Gaia~DR3~4615070113356341632 observed in Sector~65, illustrating the successive steps of the cleaning procedure.
(a) Raw TESS flux measurements as extracted from the SAP light curve.
(b) Light curve after detrending via linear regression to remove long-term instrumental and systematic trends.
(c) Detrended light curve binned to a 30-minute cadence to reduce high-frequency noise.
(d) Final cleaned light curve after removal of outliers, which is used for all subsequent variability and rotation-period analyses.
}
\label{fig:detrend}
\end{figure}

To enable a direct comparison between our automated rotation detection pipeline and published rotation rates, we used the same TESS sectors as those in the studies cited in Section~\ref{sec:data} (Tucana--Horologium Association, IC~2602: Sectors 1–65; Pisces--Eridanus: Sectors 1–5; Group~X: Sectors 1–25). Prior to any modeling, we apply a uniform and fully automated pre-processing pipeline to all light curves in order to ensure a homogeneous treatment across targets, observing sectors, and photometric pipelines. This pre-processing step includes:

\begin{itemize}

\item \textbf{Pipelines:} Our pipeline is build for the best comparison possible with literature data. For all clusters, we use the CPM and SAP pipeline. While \cite{Popinchalk2024} establishes that CPM is better suited as a detrending method for stars ($G >10$ mag) and SAP for stars with ($G < 10$ mag), that work also mentioned that stars with $9.5 \leq G \leq 10.5$ mag could be analyzed using either method, depending on the case. 

\item \textbf{Binned data:}  Each TESS FFI sector was observed with a different cadence. For instance, TESS Sectors 1--26 were observed with a 30-minute cadence, while sectors 26--52 were observed with a 10-minute cadence; sectors 53 and later were observed with a 200-second cadence.\footnote{\url{https://heasarc.gsfc.nasa.gov/docs/tess/data-products.html}} All light curves were binned to a common 30-minute cadence for our analysis, in order to minimize the impact of different cadences on the measured rotation periods.

\item \textbf{Linear fit:} We perform a linear fit on each half of the light curve of each sector to remove any overall trend.

\item \textbf{Outlier detection:} We remove an initial batch of outliers based on amplitude of the data. We remove outliers in amplitude $A$ by excluding points that lie more than 1.5 times the interquartile range beyond the 5\ith\ and 95\ith\ percentile limits, or:

\begin{equation*}
    1.5\cdot P_{5^{th}}(A)\leq A 
  \end{equation*}  
  \begin{equation*}
    A \leq 1.5 \cdot P_{95^{th}}(A)
\end{equation*}

\end{itemize}
The resulting light curves represent a consistently detrended and quality-controlled dataset, optimized for the recovery of rotational modulation while mitigating instrumental systematics and long-term trends, as seen in Figure \ref{fig:detrend}. These processed data serve as the input to the Gaussian Process modeling framework described in the next Section \ref{sec:results}.

\section{Results}\label{sec:results}
%%To correct

In this section, we assess the recovery rate of our entropy regularization method using the open clusters described above each benefiting from a number of literature TESS rotation periods. Our analysis pipeline consists of four primary stages:

\textbf{$1.$ Sample Selection.} For each cluster, we limit our analysis to cluster members with published rotation periods. The clusters in question are defined in Section \ref{sec:data}.

\textbf{$2.$ Contamination Filtering.} To reduce flux contamination, we excluded all stars located within three TESS pixels of one another. Contamination from older field stars is typically negligible due to their longer rotation timescales, which fall outside the sensitivity range of TESS photometric baselines \citep{Popinchalk2023}. However, contamination from nearby young, comoving stars, rotating on similar timescales, is nontrivial. For these cases, we recommend manual inspection, as disentangling overlapping signals remains challenging for standardized Gaussian Process (GP) models.

\textbf{$3.$ Detrending and Quality Control.} Light curves were detrended on a per-sector basis, following the procedure detailed in Section~\ref{subsec:phot}. Sectors exhibiting high variability, defined as a point-to-point root mean square (\( p2p_{\mathrm{rms}} > 0.01 \)), were automatically flagged as \texttt{high\_rms}. Similar to the analysis made in \cite{Popinchalk2023}, we interpret such large variances as indicative of instrumental artifacts or data quality issues, rather than astrophysical variability from stellar spots. For consistency with the original analyses, no alternative detrending was attempted for sectors flagged as \texttt{high\_rms}.Instead, these sectors were excluded from further analysis.

\textbf{$4.$ Rotation Modeling via Gaussian Processes.} We modeled each light curve using a GP with a quasi-periodic covariance kernel (\texttt{RotationTerm}) implemented in \texttt{celerite2}, with priors specified in Table~\ref{tab:priors}. To investigate the effect of entropy-based regularization, we evaluated performance across five values of the entropy weight \(\lambda \in \{1, 2, 4, 8, 16\}\), where \(\lambda = 1\) corresponds to the standard GP likelihood (Equation~\ref{eq:like}).

\textbf{$5.$ Rotation Detection Criteria.} Rather than adopting a set of hard thresholds to classify rotation detections, we employ a probabilistic framework in which multiple diagnostic quantities are combined into a continuous rotation likelihood. For each light-curve sector, we compute a set of complementary metrics sensitive to periodic variability, model robustness, and photometric quality. Each diagnostic probes a distinct failure mode commonly encountered in rotation period inference (e.g. aliasing, spot evolution, or instrumental systematics), and no single criterion is sufficient on its own. By combining them probabilistically, we obtain a continuous measure of confidence rather than a binary classification. For each sector, we compute the following diagnostic quantities:
\begin{enumerate}
    \item \textit{Fractional period uncertainty}.  
    The relative uncertainty of the Gaussian Process (GP)–inferred rotation period,
    \begin{equation}
        \frac{\sigma_{P}}{P} = \frac{\sigma_{+} + \sigma_{-}}{P},
    \end{equation}
    where $\sigma_{+}$ and $\sigma_{-}$ denote the upper and lower uncertainties on the period. Well-constrained rotation signals are expected to have fractional uncertainties below $\sim$10--20\%, while poorly constrained or spurious detections typically exhibit much broader posteriors.
    \item \textit{Posterior asymmetry}.  
    To quantify deviations from symmetric, well-behaved period posteriors, we define an asymmetry metric
    \begin{equation}
        A = \frac{|\sigma_{+} - \sigma_{-}|}{\sigma_{+} + \sigma_{-}}.
    \end{equation}
    Strongly asymmetric posteriors often indicate aliasing, insufficient phase coverage, or model degeneracies, and are therefore downweighted. 
    \item \textit{Lomb--Scargle periodogram support}.  
    We compute the Lomb--Scargle periodogram of the detrended light curve and measure the maximum power within a $\pm15\%$ window centered on the GP-inferred period. This metric provides independent confirmation of the detected periodicity and guards against GP fits driven primarily by correlated noise. 
    \item \textit{Signal-to-noise ratio of the modulation}.  
    The signal-to-noise ratio (SNR) of the rotational modulation is computed following standard time-domain definitions. Low-SNR signals are more susceptible to contamination from stochastic variability or instrumental systematics, particularly in short-baseline observations. 
    \item \textit{Photometric scatter}.  
    The root-mean-square (RMS) of the flux differences provides a global measure of light-curve stability. Elevated RMS values may indicate contamination, unresolved blends, or strong non-periodic variability, all of which reduce confidence in a rotational interpretation. It's important to mention that our pipeine flags \texttt{high$\_$rms} all target with an average point-to-point RMS $>0.01$.
\end{enumerate}

For each diagnostic, the quantity $x$ corresponds directly to the metric defined above. Specifically, $x$ is taken to be the fractional period uncertainty ($\sigma_P/P$), the posterior asymmetry ($A$), the maximum Lomb--Scargle power within the $\pm15\%$ search window, the SNR, or the point-to-point RMS scatter of the light curve. These quantities are chosen to probe complementary aspects of rotation period reliability, including the precision of the inferred period, the shape of the posterior distribution, independent periodogram support, signal significance, and overall photometric stability.
Each diagnostic quantity $x$ is mapped onto a probability-like score using a logistic (sigmoid) function,
\begin{equation}
    P(x) = \frac{1}{1 + \exp\left[-(x - x_0)/s\right]},
\end{equation}
where $x_0$ represents a characteristic transition value motivated by empirical and literature-based expectations, and $s$ is a scale parameter that controls the smoothness of the transition. The adopted values for each diagnostic (Table~\ref{tab:sigmoid_params}) were chosen empirically based on the observed distributions of the diagnostics in our sample, guided by literature expectations for reliable rotation detections. In practice, $x_0$ sets the approximate boundary between likely and unreliable detections, while $s$ determines how gradually the corresponding probability changes across that boundary. This formulation replaces abrupt pass/fail criteria with a continuous weighting scheme that better reflects the intrinsic uncertainty in rotation detection.

The individual probability terms are then combined by taking their mean to yield a sector-level rotation probability. This approach ensures that strong agreement across diagnostics produces a high-confidence detection, while marginal or inconsistent evidence is naturally downweighted rather than discarded.

For stars observed in multiple sectors, sector-level probabilities are aggregated at the star level by computing the mean rotation probability across all available sectors, as described in Equation~\ref{eq:succrate}. A star is classified as a rotator if this mean probability exceeds 0.7. The reported rotation period for each star corresponds to the sector with the highest individual rotation probability, ensuring that the quoted period reflects the most reliable detection while the overall classification incorporates consistency across observations.

\begin{table}
\caption{Diagnostic criteria used in the probabilistic rotation classifier and their corresponding sigmoid parameters.}
\label{tab:sigmoid_params}
\begin{tabular}{lcccc}
\hline
Criterion & $x_0$ & $s$ \\
\hline
Fractional period uncertainty &
0.20 &
0.05 \\

Posterior asymmetry &
0.30 &
0.05 
\\

Lomb--Scargle support &
0.30 &
0.10 \\

Signal-to-noise ratio &
10 &
3  \\

Photometric scatter &
0.01 &
0.002 \\
\hline
\end{tabular}
\end{table}

\subsection{Benchmark Period Recovery}

We quantify the benchmark period recovery using a success metric defined as:
\begin{equation}\label{eq:succrate}
\text{success rate} = Crit\,\, A \times  Crit\,\, B,
\end{equation}
where $Crit\,\, A$ is the fraction of stars classified as rotators under the probabilistic criterion, and $Crit\,\, B$ is the fraction of these rotators whose inferred rotation periods agree with published literature values to within 15\%.

\begin{figure*}[ht!] 
\plotone{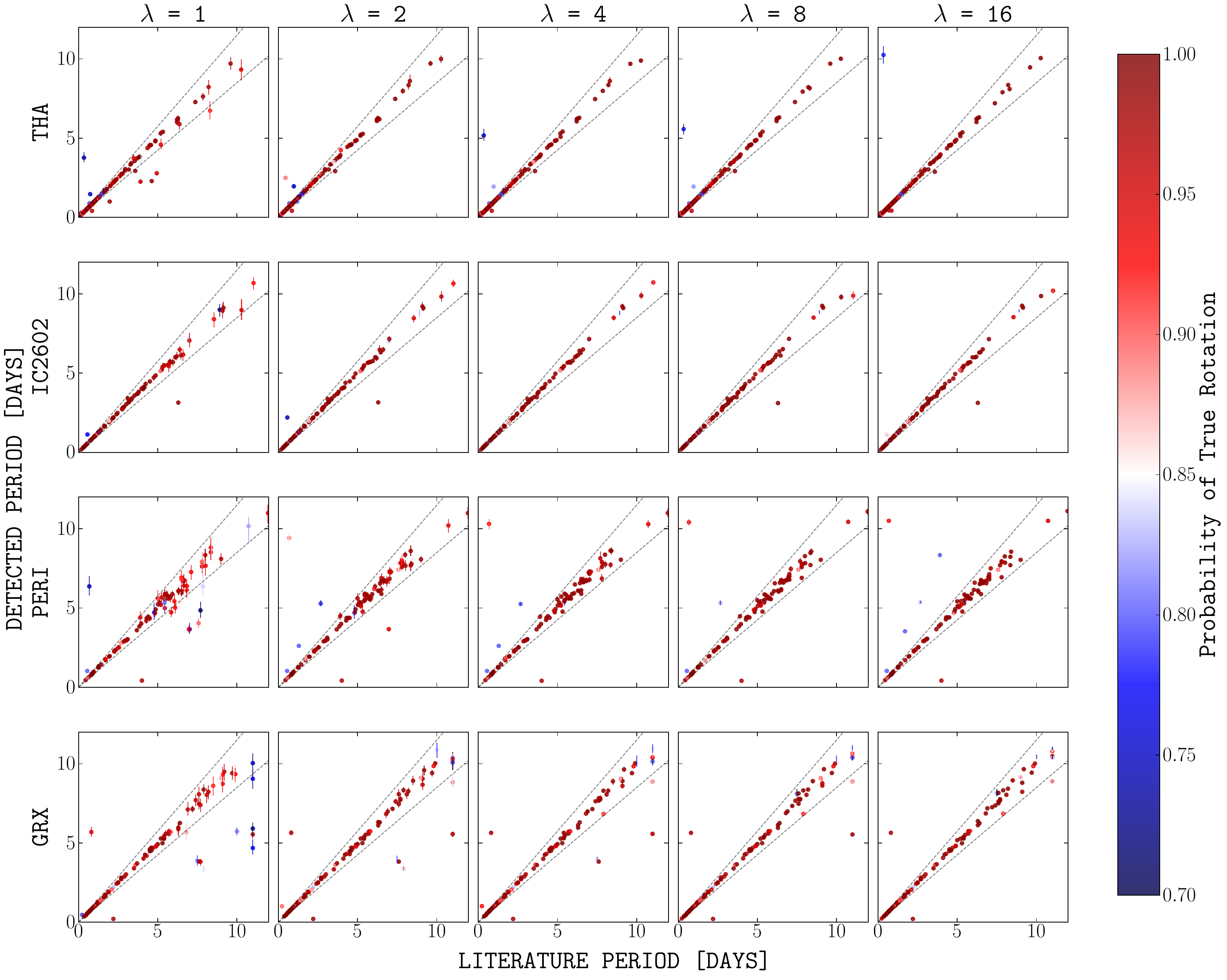} 
\caption{
Comparison of literature rotation periods with the rotation periods detected for stars Tucana--Horologium Association, IC~2602, Pisces--Eridanus and Group~X. Dashed bars indicate the 15\% range around the literature periods. The color scale represents the combined probabilistic criterion for stars with $\mathrm{prob} > 0.7$. The detailed recovery rates are shown in Table~\ref{tab:lambda}.}\label{fig:diagplot}
\end{figure*}

Figure~\ref{fig:diagplot} compares the inferred rotation periods to literature values for stars satisfying the probabilistic rotation criterion across all clusters (Tucana--Horologium Association, IC~2602, Pisces--Eridanus, and Group~X) and for a range of regularization strengths, $\lambda = 1, 2, 4, 8,$ and $16$. We find that the majority of selected rotators cluster tightly around the one-to-one relation, with no strong systematic deviations across the range of periods probed. Stars failing the 15\% agreement criterion predominantly exhibit lower rotation probabilities or inconsistent sector-level detections, supporting the effectiveness of the probabilistic framework in downweighting ambiguous cases.

\begin{figure*}[ht!]
\plotone{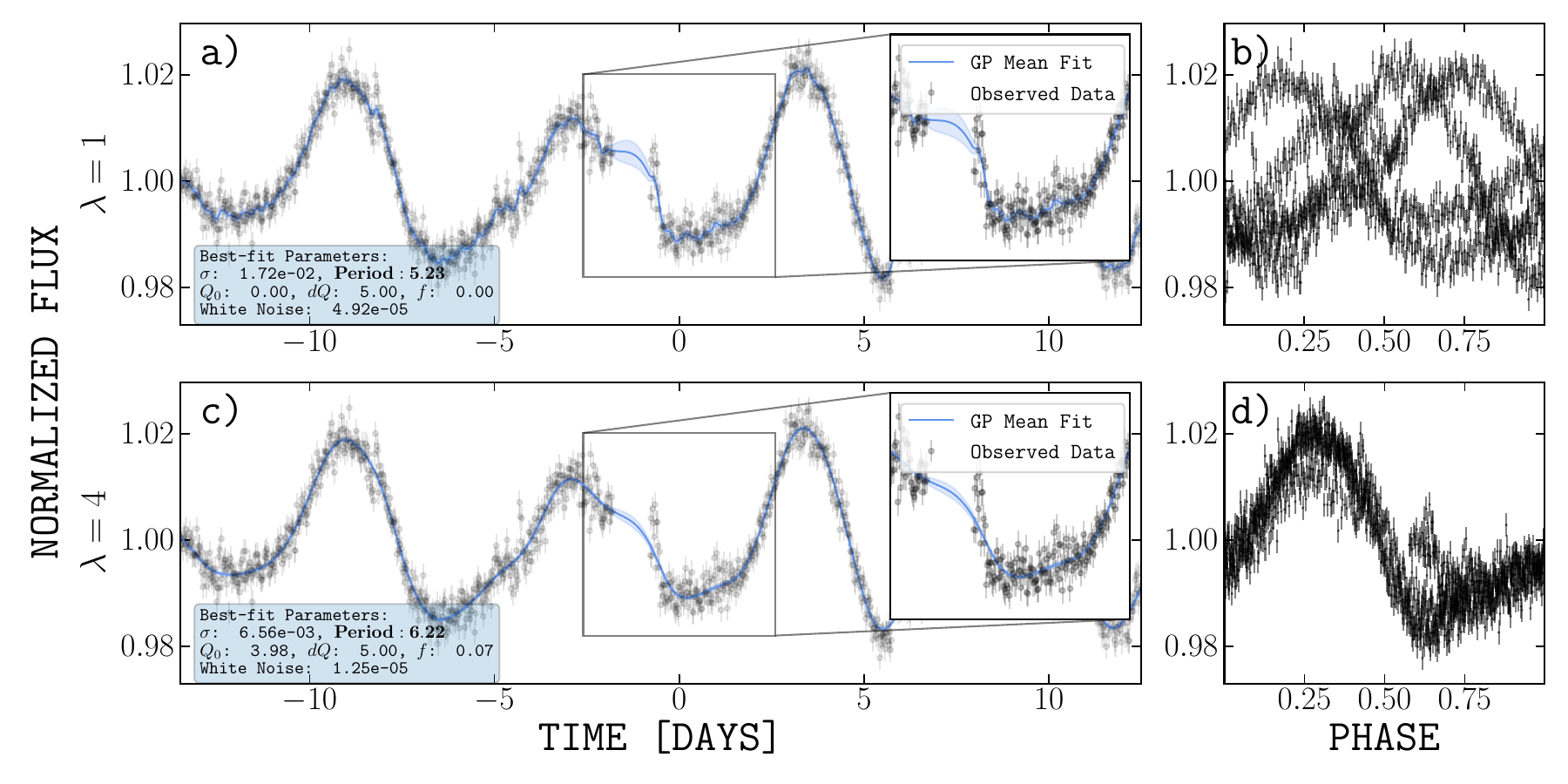}
\caption{
TESS Sector~32 light curve of Gaia DR3~4784114803446931200, a member of the Tucana--Horologium Association, modeled using GPs. \textit{Top row:} \textbf{a)} GP fit with $\lambda= 1$ (no regularization). The left panel shows the full light curve with the GP mean model and predictive uncertainty overplotted; \textbf{b)} shows the corresponding phase-folded light curve using the period inferred by the GP. \textit{Bottom row:} \textbf{c)} GP fit with $\lambda = 4$; \textbf{d)} shows the corresponding phase-folded light curve using the period inferred by the GP. Increasing $\lambda$ yields a smoother GP model and recovers a rotation period that is more physically consistent, as evidenced by the improved coherence of the phase-folded light curve and consistency with the literature period of 6.22 days.} \label{fig:gpfit}
\end{figure*}

Increasing the regularization strength $\lambda$ generally improves the robustness of rotation period recovery by suppressing overfitting and reducing spurious detections (see Figure~\ref{fig:gpfit}). However, excessive regularization can lead to overly simplistic models that fail to capture rotational modulation, highlighting the importance of exploring a range of $\lambda$ values. As shown in Table~\ref{tab:lambda}, the optimal value of $\lambda$ varies between clusters, although moderate regularization ($\lambda = 2$–16) consistently yields recovery rates comparable to or better than the unregularized case.

The combination of GP regularization and probabilistic classification yields improved agreement with literature periods relative to binary selection criteria. All diagnostic plots and validation products are publicly available at \texttt{mocadb}.\footnote{\url{https://dataviz.mocadb.ca/moca-explorer}} These diagnostics also enabled us to revise rotation periods for select targets based on improved fits, which will be further described in Section \ref{subsec:revise}.

\newpage

\tablewidth{1\textwidth}
\begin{deluxetable}{cccc}[!ht] \label{tab:lambda}
\tablecolumns{4}
\tablecaption{Recovery rate for rotation detection across various penalty.}
\tablehead{
\colhead{ Penalty Weight} &
\colhead{$Crit\,\,A$} &
\colhead{$Crit\,\,B$} &
\colhead{\textbf{Success Rate}}}
\startdata
\cutinhead{{Tucana-Horologium Association}}
$\lambda = 1$ & 97.78\% & 94.57\% & $\textbf{92.48\%}$  \\
$\lambda = 2$ & 97.78\% & 97.29\% & $\textbf{95.13\%}$   \\
$\lambda = 4$ & 99.56\% & 96.89\% & $\textbf{96.46\%}$   \\
$\lambda = 8$ & 99.56\% & 96.89\% & $\textbf{96.46\%}$   \\
$\lambda = 16$ & 99.56\% & 97.33\% & $\boxed{\textbf{96.90\%}}$   \\
\cutinhead{{IC\,2602}}
$\lambda = 1$ & 100.00\% & 98.04\% & $\textbf{98.04\%}$   \\
$\lambda = 2$ & 100.00\% & 98.04\% & $\textbf{98.04\%}$   \\  
$\lambda = 4$ & 100.00\% & 100.0\% & $\boxed{\textbf{100.00\%}}$   \\
$\lambda = 8$ & 100.00\% & 99.02\% & $\textbf{99.02\%}$   \\
$\lambda = 16$ & 100.00\% & 98.03\% & $\textbf{98.04\%}$   \\
\cutinhead{{Pisces-Eridanus}}
$\lambda = 1$ & 94.89\% & 86.02\% & $\textbf{81.63\%}$   \\
$\lambda = 2$ & 98.98\% & 91.75\% & $\textbf{90.82\%}$   \\
$\lambda = 4$ & 98.98\% & 93.81\% & $\textbf{92.86\%}$   \\
$\lambda = 8$ & 98.98\% & 94.85\% & $\boxed{\textbf{93.88\%}}$   \\
$\lambda = 16$ & 98.98\% & 92.78\% & $\textbf{91.83\%}$   \\
\cutinhead{{Group X} }
$\lambda = 1$ & 99.10\% & 88.18\% & $\textbf{87.39\%}$   \\
$\lambda = 2$ & 100.00\% & 92.79\% & $\textbf{92.79\%}$   \\
$\lambda = 4$ & 100.00\% & 93.69\% & $\textbf{93.69\%}$   \\
$\lambda = 8$ & 100.00\% & 96.40\% & $\textbf{96.40\%}$   \\
$\lambda = 16$ & 100.00\% & 97.30\% & $\boxed{\textbf{97.30\%}}$   \\
\enddata
\end{deluxetable}

%% Please use the acknowledgment and contribution environments. This will 
%% be anonomyized when the "anonymous" style option is used. 

\section{Discussion}\label{sec:discussion}
In this section, we assess the efficacy of regularized Gaussian Processes in recovering stellar rotation periods from high-precision TESS photometry. Our results demonstrate that introducing a regularization factor, $\lambda$, improves the physical interpretability of the GP hyperparameters by rebalancing the likelihood’s data-fit and complexity terms. Across all clusters and for a range of regularization strengths ($\lambda = 1, 2, 4, 8,$ and $16$), we find that moderate regularization yields the most robust period recovery, as seen in Figures~\ref{fig:diagplot} and \ref{fig:gpfit}.

At low regularization ($\lambda = 1$), the model tends to overfit the data, leading to increased scatter and ambiguous period detections. In contrast, very high values of $\lambda$ (e.g., $\lambda = 16$) can oversimplify the covariance structure, suppressing genuine rotational modulation. This tradeoff highlights the importance of exploring a range of $\lambda$ values rather than adopting a single fixed choice. As summarized in Table~\ref{tab:lambda}, the optimal value of $\lambda$ varies between clusters, although moderate values ($\lambda \sim 2$–16) generally provide the best balance between robustness and flexibility.

Interestingly, the sensitivity of the inferred rotation period to variations in $\lambda$ serves as a diagnostic tool in itself. Consistency in period recovery across multiple regularization strengths typically indicates a robust, physically-driven signal. Conversely, cases where the detected periodicity shifts or vanishes as $\lambda$ increases often point to the presence of instrumental systematics, complex spot evolution, or other non-stationary behaviors that a standard GP might otherwise ``overfit" as a clean rotation signal. These edge cases highlight both the strengths and limitations of our current approach and provide physical insight into the diverse stellar variability observed in young populations.

Despite the robustness introduced by hyperparameter regularization, certain light-curve behaviors continue to elude automated period recovery. We detail these persistent failure modes below, and provide in Table~\ref{tab:rot_all} the detected rotation rate for all clusters, pipeline and values of $\lambda$.

\begin{deluxetable*}{rlll}
\tablecaption{Machine-readable Rotation Catalog Column Descriptions\label{tab:rot_all}}
\tablehead{
\colhead{Column} & \colhead{Label} & \colhead{Units} & \colhead{Description}
}
\startdata
1 & GaiaDR3 & \nodata & Gaia DR3 source identifier \\
2 & Assoc & \nodata & Stellar association or cluster \\
3 & Sector & \nodata & TESS sector number \\
4 & Pipeline & \nodata & Light-curve reduction pipeline \\
5 & M & \nodata & Gaussian-process regularization factor \\
6 & Plit & d & Literature rotation period \\
7 & Category & \nodata & Rotation classification category \\
8 & Gmag & mag & Gaia DR3 mean G-band magnitude \\
9 & G-RP & mag & Gaia DR3 G-RP color \\
10 & BP-RP & mag & Gaia DR3 BP-RP color \\
11 & Prot & d & Gaussian-process rotation period, rounded to the decimal place set by its uncertainties \\
12 & e\_Prot & d & Lower uncertainty on Prot, quoted to two significant figures\tablenotemark{a} \\
13 & E\_Prot & d & Upper uncertainty on Prot, quoted to two significant figures\tablenotemark{a} \\
14 & pRot & \nodata & Rotation-detection probability \\
15 & pSNR & \nodata & Signal-to-noise probability \\
16 & pAsym & \nodata & Light-curve-asymmetry probability \\
17 & pRMS & \nodata & Light-curve-RMS probability \\
18 & pLS & \nodata & Lomb-Scargle probability \\
19 & pFinal & \nodata & Final combined rotation probability \\
20 & Sel & \nodata & Adopted-period flag: 1=selected, 0=not selected\tablenotemark{b} \\
\enddata
\tablenotetext{a}{e\_Prot and E\_Prot give the lower and upper uncertainties, respectively, on the Gaussian-process rotation period, corresponding to the differences between the adopted period and the 16th and 84th percentiles of the marginalized posterior distribution.}
\tablenotetext{b}{A value of 1 identifies the adopted rotation-period measurement; 0 indicates a non-adopted row.}
\tablecomments{The complete table is published in its entirety in machine-readable format. This descriptive table is provided for guidance regarding its form and content.}
\end{deluxetable*}

\paragraph{\textbf{Case A: Multiple Periodicities}}

A subset of targets exhibits pronounced beat-like patterns in their light curves, characterized by the superposition of multiple periodic signals with comparable amplitudes. While such behavior can arise from photometric contamination by unresolved companions within a single TESS pixel, this explanation is unlikely for our sample, as we explicitly exclude close stellar pairs and known contaminants (Section~\ref{sec:results}).
More plausible interpretations involve stellar differential rotation, where starspots at different latitudes rotate at distinct angular velocities, imprinting multiple periodicities in the photometric signal (e.g., \citealp{Reinhold2013}), or even unresolved binaries, where the both stars have different rotation rates, causing multiple rotation signals in a single light curve. In other cases, such as for Gaia~DR3~51145160203690388, for which we revised the rotation rate from the literature value, the light curve could simply be contaminated by an unresolved young moving group candidate that wasn't found yet. In such cases, a single-period GP model may struggle to converge on a unique rotation period or may recover only one of the rotation signals.

\begin{figure}[ht]
\includegraphics[width=8.5cm]{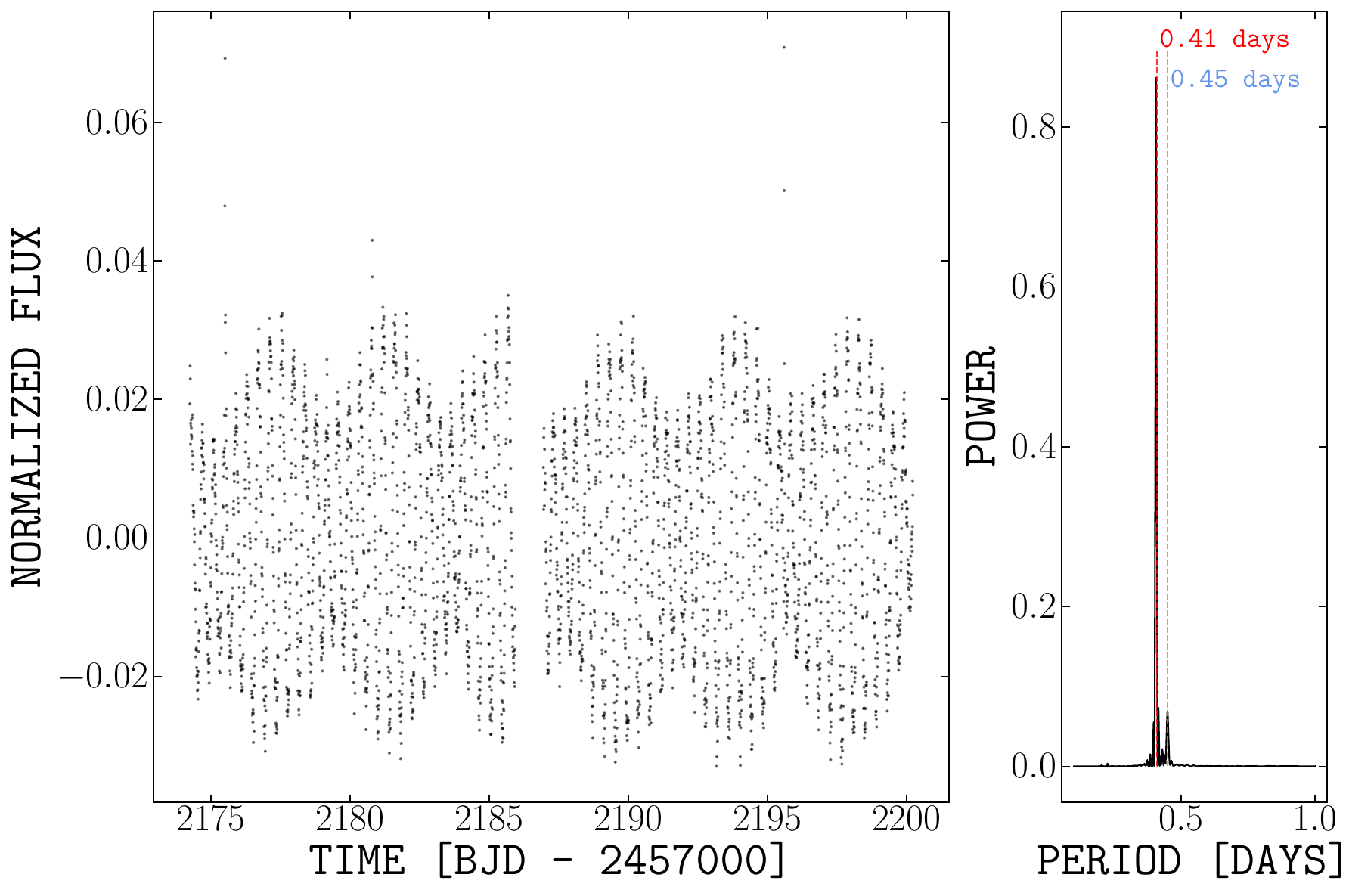}
\caption{ \textbf{Case A: Beat Patterns} --
\textit{Left:} Light curve of Gaia~DR3~3177883999240571904 observed in TESS Sector 66, extracted using the CPM pipeline. 
\textit{Middle:} Lomb--Scargle periodogram derived from the same light curve. While the periodogram exhibits 2 clear rotational peak, at both 0.41 and 0.45 day periods, hinting at unresolved binaries.}
\label{fig:caseA}
\end{figure}

An illustrative example, as seen in Figure \ref{fig:caseA} is the Pisces--Eridanus target Gaia~DR3~3177883999240571904, whose light curve displays a clear beat pattern. Across all available sectors, our GP model consistently converges on a short rotation period of 0.41~days, whereas \citet{Curtis2019} adopted a longer period of approximately 4~days. Both signals are present in the data as well as a second short period of 0.45 days when we inspect the periodogram as  seen in Figure \ref{fig:caseA}. A detailed analysis of this target enable us to hypothesize the presence of a likely unresolved binary, which would explain the two short periods causing the larger beat pattern. This object has a Gaia Renormalized Unit Weight Error (RUWE) of 1.53, pointing to a poor a single-star model astrometric fit. Moreover, isochronal tracks seem to suggest that given the age of Pisces-Eridanus ($\sim$120 Myr), the object is $\approx 2$\,mag brighter than expected.

Another example from Group~X of a light curve showcasing more than one rotation rates is  Gaia~DR3~1592384486775784064. With a RUWE of 7.98636, this object is most likely an unresolved binary, which explains the multiple signal seen in the light curve across all sectors.

\paragraph{\textbf{Case B: Time-Variable Amplitude}} Another common failure mode arises in high signal-to-noise targets, where the rotational modulation is clearly detected, but its amplitude varies significantly over the duration of the observations. Here, we define significant amplitude evolution as changes in the peak-to-peak modulation amplitude of order tens of percent across adjacent or nearby rotation cycles. While young, magnetically active stars commonly show evolving spot patterns, smoothly repeating light curves are generally expected over a few rotation periods. Rapid amplitude evolution can therefore indicate spot emergence, migration, or decay on timescales comparable to the rotation period.

When amplitude variations are strong, the GP model may struggle to identify a stable set of hyperparameters, particularly those governing the rotation period and the amplitude-related terms (e.g., $\sigma$). This can lead to broad or multimodal posterior distributions, or, in some cases, non-convergent solutions. While our regularized GP framework improves robustness, fully capturing such behavior may require more complex models, such as mixtures of quasi-periodic kernels or explicitly time-dependent amplitudes.

\begin{figure*}[ht]
\plotone{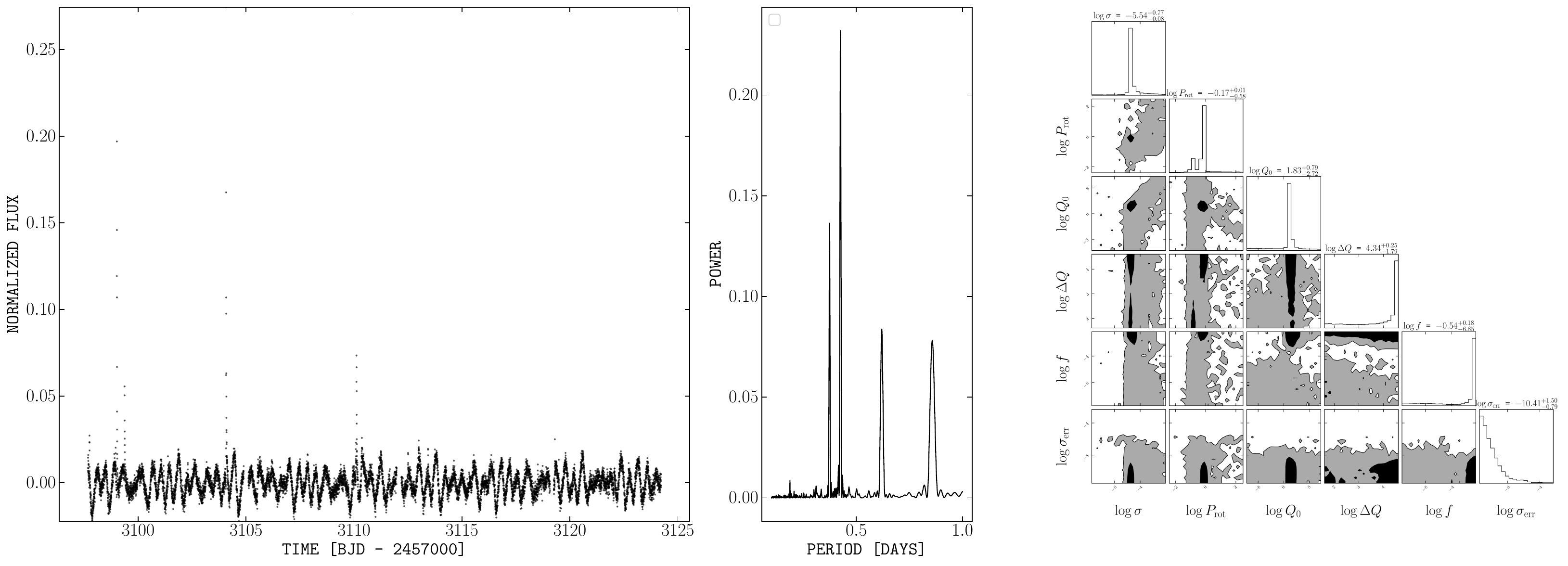}
\caption{ \textbf{Case B: Time-Variable Amplitude  } --
\textit{Left:} Light curve of Gaia~DR3~5764784514344281088 observed in TESS Sector 66, extracted using the CPM pipeline. 
\textit{Middle:} Lomb--Scargle periodogram derived from the same light curve. While the periodogram exhibits clear rotational peaks, 
the photometric amplitude varies significantly from rotation to rotation.
\textit{Right:} Corner plot of the GP hyperparameter, which showcases the lack of robust convergence of the model for $\lambda = 4$.
}
\label{fig:caseb}
\end{figure*}

Several targets exemplify this behavior. In THA, Gaia~DR3~5764784514344281088 exhibits a complex combination of uneven modulations, beat-like features, and sector-to-sector inconsistencies in the inferred rotation period, making it an especially compelling case for detailed follow-up. The specific case is illustrated in Figure~\ref{fig:caseb}.

In PERI, Gaia~DR3~2492898356897645184 shows strong amplitude variability, particularly in Sector~31, leading our pipeline to favor a period twice that reported in the literature. Sector~4 further displays signs of imperfect de-trending mid-sector, compounding the ambiguity.

Similarly, Gaia~DR3~5182223980765557248 is inferred to have a rotation period approximately twice the literature value. This behavior likely reflects the structure of the GP \texttt{RotationTerm}, which models variability as a combination of a fundamental period and its harmonic. If the amplitude varies substantially from one rotation cycle to the next, the model may interpret this as evidence for a harmonic component and converge on a longer effective period. A related effect is seen in Gaia~DR3~3185678437170300800, where the light curve appears modulated by a slower envelope ($\sim$6.16~days) that induces apparent amplitude variations not explicitly modeled in our framework.

\paragraph{\textbf{Case C: Bad Detrending}}

In a smaller number of cases, failures can be traced primarily to imperfect de-trending of the light curves, resulting in residual long-term trends or instrumental systematics that obscure the rotational signal. Such effects can suppress or distort periodic variability, particularly when only a single sector is available for analysis.

\begin{figure}[ht]
\includegraphics[width=8.5cm]{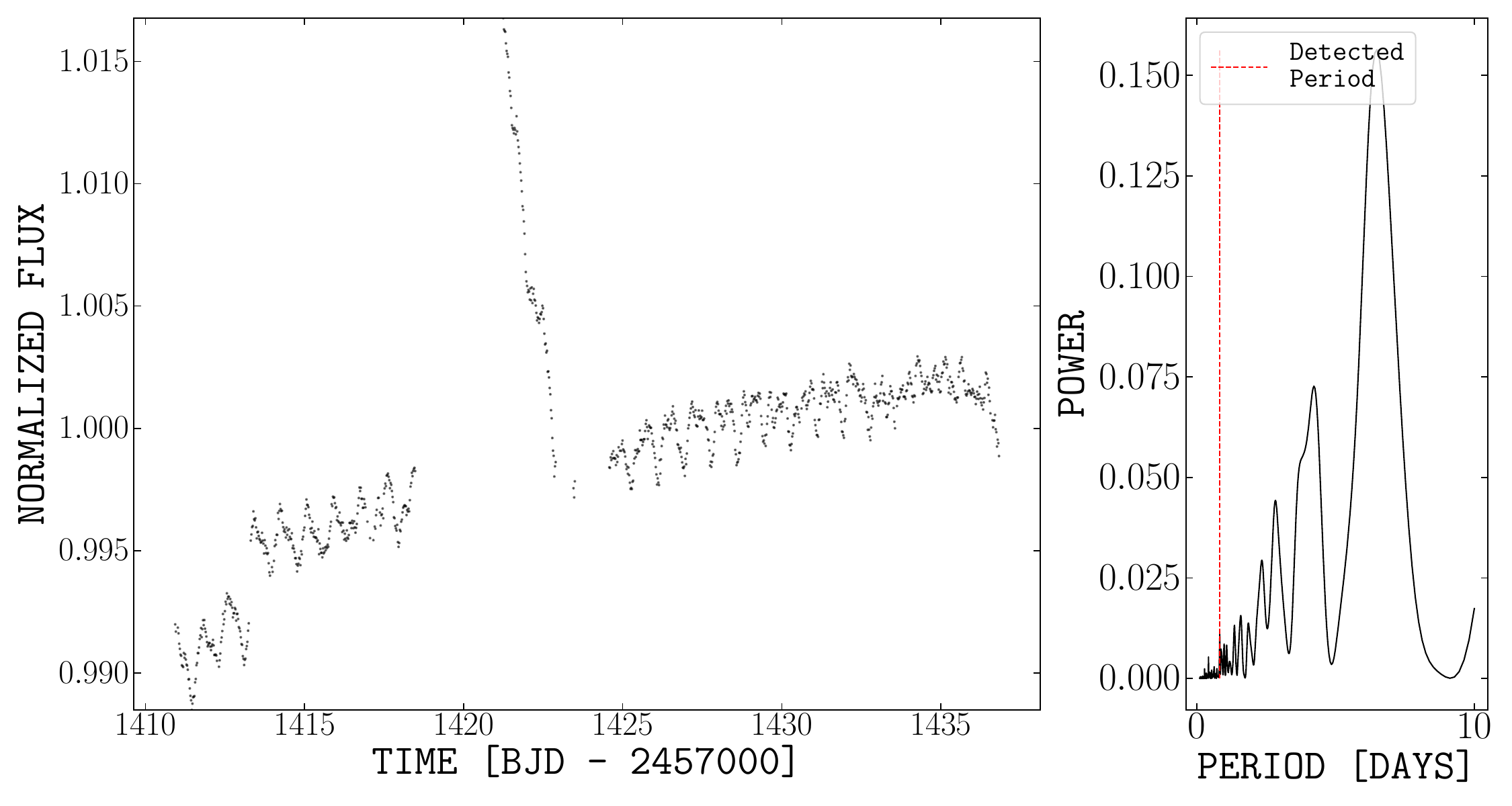}
\caption{ \textbf{Case C: Bad Detrending  } --
\textit{Left:} Light curve of Gaia~DR3~2516948215250061568 observed in TESS Sector 4, extracted using the SAP pipeline. The light curve shows non-linear systematics, caused by the data reduction.
\textit{Right:} Lomb--Scargle periodogram derived from the same light curve.
}
\label{fig:casec}
\end{figure}

An example in the Tucana--Horologium Association is Gaia~DR3~4936685751336237056, where residual trends dominate the light curve, as seen in Figure~\ref{fig:casec}. In PERI, Gaia~DR3~2516948215250061568 shows significant systematics in Sectors~4 and~31, while later sectors (42, 43, 70, and 71) yield rotation periods consistent with those reported by \cite{Curtis2019}. To be consistent with \cite{Curtis2019}, we only allowed our Sector~4 to be used for the automated rotation detection.
Moreover, Gaia~DR3~1404937579808394112, from Group~X, presents a similar case, where all sectors (23, 24, 49, 50, 51, 76, 77, 78). Because the target is bright ($G_{mag}=8.94$), the poor SAP detrending is unlikely to result from low SNR. The behavior more likely reflects saturation-related systematics.

The importance of a good, systematic-free light curve for proper GP convergence cannot be stressed enough.
It's important to note that this case could be fixed by the use of a more complex kernel, such as the combination of a kernel that models the stellar rotation, and another that models systematics.

\paragraph{\textbf{Case D: Miscellaneous and Ambiguous Cases}}
A small number of targets defy aforementioned classification. For example, the Tucana--Horologium Association member Gaia~DR3~4843959091042621312 shows no significant Lomb--Scargle power at the inferred GP period of 0.89\,days.

\begin{figure*}[ht]
\plotone{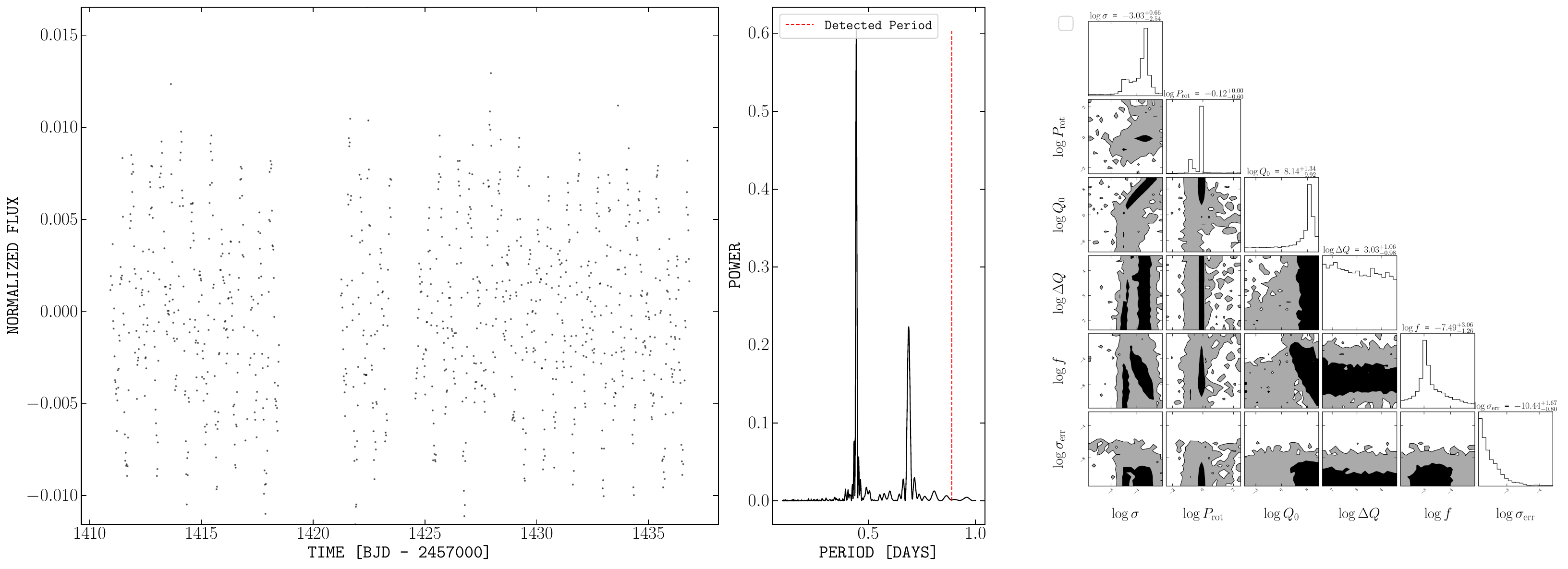}
\caption{ \textbf{Case D: Miscellaneous Cases } --
\textit{Left:} Light curve of Gaia~DR3~4843959091042621312 observed in TESS Sector 4, extracted using the CPM pipeline. 
\textit{Middle:} Lomb--Scargle periodogram derived from the same light curve. While the periodogram exhibits clear rotational peaks, at 0.44 and 0.68 days, the GP seems to converge strongly towards a 0.89-day peak.
\textit{Right:} Corner plot of the GP hyperparameter for $\lambda = 4$. For the period hyperparameters (Log $P_{rot}$), we see that the model explored both $P_{rot} = 0.44\,d$ and $P_{rot} = 0.89\,d$.
}
\label{fig:cased1}
\end{figure*}

The apparent convergence likely arises from the interplay between multiple periodic signals present in the light curve, as seen in Figure~\ref{fig:cased1}. While the Lomb--Scargle periodogram exhibits strong peaks at 0.44 and 0.68 days, the GP converges to a period near 0.89 days, which is approximately twice the dominant 0.44-day signal. Because our GP kernel explicitly accounts for both a fundamental period and its first harmonic ($P_{\rm rot}/2$), the model naturally explores solutions near both 0.44 and 0.89 days. Inspection of the posterior distribution of $\log P_{\rm rot}$ indicates that the sampler extensively explored the 0.44-day solution but failed to converge there, likely because the additional 0.68-day variability distorted the covariance structure and favored the longer-period interpretation. With stronger regularization ($\lambda=16$), however, the model ultimately converges to the 0.44-day signal, suggesting that the 0.89-day solution is a consequence of model degeneracy between the harmonic and multi-periodic components rather than an independently supported rotation period.

Another example is Gaia~DR3~1413990168277612928, from Group~X, which simply has a low SNR and a long rotation period. The weak modulation and limited phase coverage available within a TESS sector make the rotational signal difficult to distinguish from noise, preventing a robust recovery of the literature period.

\subsection{Revision of Literature Rotation Periods}\label{subsec:revise}

Our analysis also identifies several targets for which the available photometric evidence suggests that the literature rotation period may be incorrect or ambiguous.

In Pisces--Eridanus, Gaia~DR3~5114516020369038848 has a reported rotation period of 0.68\,d, which is clearly detected in the SAP light curve but not in the CPM reduction. Given the brightness of the target ($G = 13.36$), CPM is expected to provide a more reliable correction of systematics and crowding effects than SAP. The absence of the signal in the CPM light curve therefore suggests that the SAP detection does not originate from the target itself.

\begin{figure*}[ht!] 
\plotone{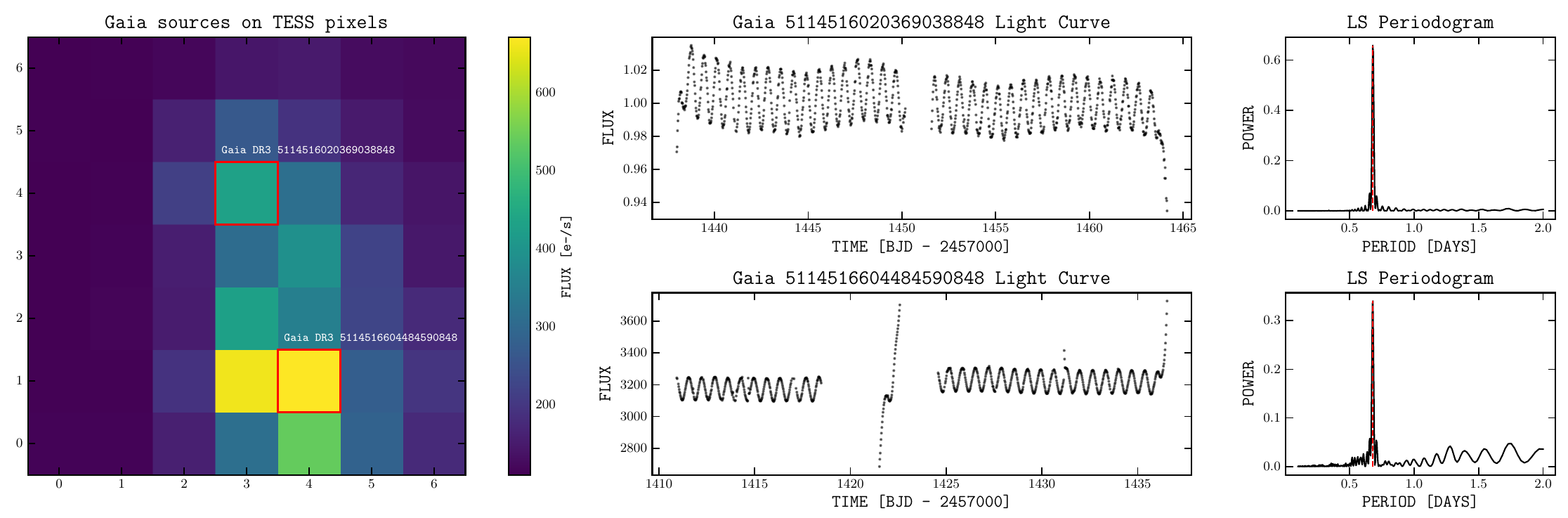} 
\caption{ \textbf{Revised period} --
TESS Target Pixel File (left) for Gaia~DR3~5114516020369038848, with Gaia DR3 sources overlaid as red squares. 
Each row on the right shows the extracted light curve (middle column) and Lomb--Scargle periodogram (right column) 
for the corresponding Gaia source in the cutout. The light curves are generated using circular apertures centered 
on the Gaia positions from the package \texttt{lightkurve}. Notably, the closest neighbor to the target shows no significant photometric modulation, 
indicating that the observed period in our target stems from its neighboring, brighter object Gaia~DR3~5114516604484590848.}
\label{fig:caseRevisedContam}
\end{figure*}

The updated membership analysis of \cite{mocapaper} identifies Gaia~DR3~5114516604484590848, a nearby and brighter star ($G = 12.20$), as a cluster member exhibiting a 0.68\,d periodicity. The spatial proximity and identical period strongly indicate that the signal detected in the SAP light curve of Gaia~DR3~5114516020369038848 is due to flux contamination from this neighbor. This interpretation is supported by the light curve comparison shown in Figure~\ref{fig:caseRevisedContam}, as well as the Gaia Variability Catalog \citep{GaiaVar2022}, which confirms that the rotation period of 0.68\,d is associated with Gaia~DR3~5114516020369038848.

In THA, Gaia~DR3~4712271339297039744 was reported by \cite{Popinchalk2023} to have a rotation period of 0.46~days. However, the light curve strongly indicates that this value corresponds to the first harmonic, with the true rotation period closer to 0.91~days. Similarly, Gaia~DR3~6388075665896356480 exhibits two periodic signals (2.35~days and $\sim$0.4~days) in early sectors analyzed by \cite{Popinchalk2023}, who favored the longer period. In later sectors (94 and 95), the 2.35-day signal disappears entirely, leaving only the shorter-period modulation. This evolution may reflect changes in dominant spot latitudes, spot lifetimes, or differential rotation, underscoring the importance of multi-sector analyses.

In Pisces--Eridanus, Gaia~DR3~2349094158814399104 was previously analyzed using only Sector~3 by \cite{Curtis2019}. While our analysis of Sector~30 recovers a strong signal near 0.67~\,d, the light curve exhibits clear multi-periodic behavior, with additional power near 0.42~\,d. The near-integer relationship between the periods (($\sim0.67$), ($\sim0.42$), and ($\sim1.3$)~\,d) further complicates the interpretation and may contribute to the apparent 1.30-day signal. Rather than representing a simple harmonic, the observed variability is more consistent with a multi-periodic system, potentially arising from unresolved binary components with distinct rotation periods. 

Likewise, for Gaia~DR3~5094664333632217088, we find that Sectors~4, ~5 and~31 are more consistent with a rotation period of 1.84~\,d than the 2.62-day value favored in the literature. This agreement is particularly notable given that our pipeline is explicitly designed to identify and penalize harmonic misidentifications.

Together, these cases illustrate how regularized GP modeling, combined with multi-sector photometry and probabilistic classification, can both recover robust rotation periods and highlight targets where existing literature values warrant re-examination. We note, however, that several of the limitations discussed above are not unique to the GP framework presented here. The specific contribution of our regularized GP framework is not to eliminate these general limitations, but to provide an additional diagnostic of whether the period inferred from a flexible covariance model remains physically interpretable across sectors and regularization strengths.

\subsection{Complex Rotators}
Complex rotators are a subclass of stars exhibiting non-sinusoidal photometric modulation, often characterized by evolving or multi-peaked light curve shapes \citep{Rebull2016,Stauffer2016,Popinchalk2023, Bouma2023}. These objects are typically rapidly rotating M dwarfs (with $P_{\rm rot} < 1$ day) and have been studied in detail by, e.g., \citet{Stauffer2017, Popinchalk2023}. While the exact physical mechanisms behind these morphologies remain debated, ranging from persistent high-latitude starspots, differential rotation, to star–disk interactions, their defining observational feature is a deviation from a clean, sinusoidal-shaped light curve.

\begin{figure}[ht]
\centering
\includegraphics[width=8.5cm]{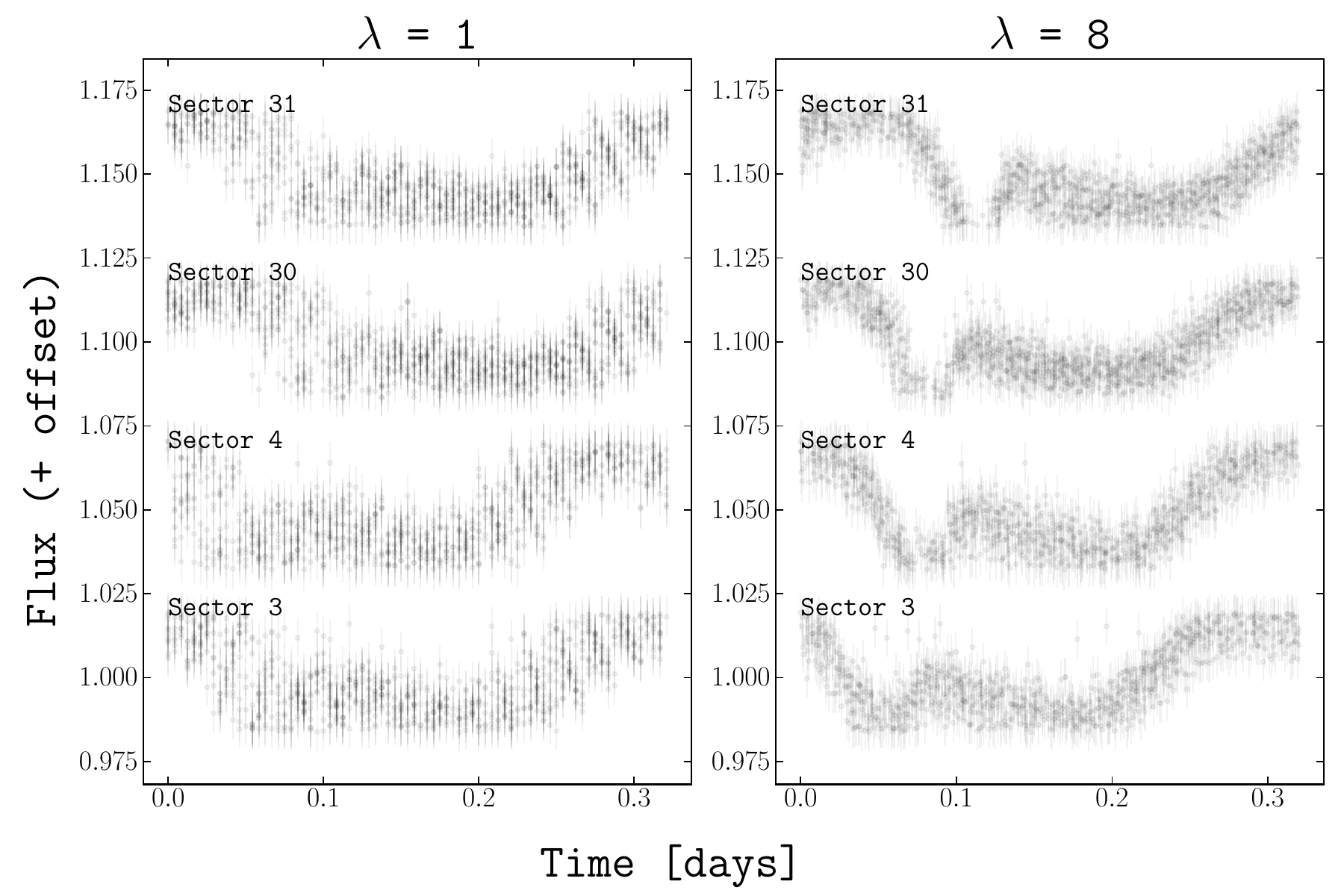}
\caption{Phase-folded light curves for each sector of a complex rotator Gaia DR3 5070527824315406976 from (THA) under increasing levels of regularization $\lambda$. \textit{Left plot:} Phase-folded light curve for the fitted rotation rate with a GP wit $\lambda = 1$ (no entropy regularization).  \textit{Right plot:} Phase-folded light curve for the fitted rotation rate with a GP wit $\lambda = 8$. For values of $\lambda = 2, 4, 8, 16$, the rotation period becomes more stable thanks to the regularization and converges toward literature values despite the non-sinusoidal morphology.}
\label{fig:complexrot}
\end{figure}

Our dataset includes four such complex rotators previously identified in \citet{Popinchalk2023, Popinchalk2024}. In Figure~\ref{fig:complexrot}, we show an example from the IC\,2602 association. The light curve is phase-folded using rotation periods inferred under varying levels of entropy regularization (i.e., different values of $\lambda$). As $\lambda$ increases, the GP solution converges more consistently to the rotation period reported in the literature, despite the light curve’s irregular morphology.

This example highlights a major strength of our approach: Gaussian Processes, when properly regularized, are capable of modeling a wide range of variability morphologies, that is not limited to sinusoidal variations. Their non-parametric nature enables them to fit complex periodic behavior while still constraining against overfitting. 

\subsection{Sector-to-Sector Consistency}
\label{sec:cross_sector}

For stars observed in multiple TESS sectors, we tested whether the inferred rotation period remained consistent across independent observations. We quantified the cross-sector variability using the median absolute deviation,

\begin{equation}
D_{\mathrm{sector}}
= 1.4826
\operatorname{median}
\left(
\frac{\left|P_s-\widetilde{P}\right|}
{\widetilde{P}}
\right),
\label{eq:sector_fractional_mad}
\end{equation}
where $P_s$ is the period inferred from sector $s$ and $\widetilde{P}$ is the median period across all usable sectors for that star.

\begin{figure*}[ht!] 
\plotone{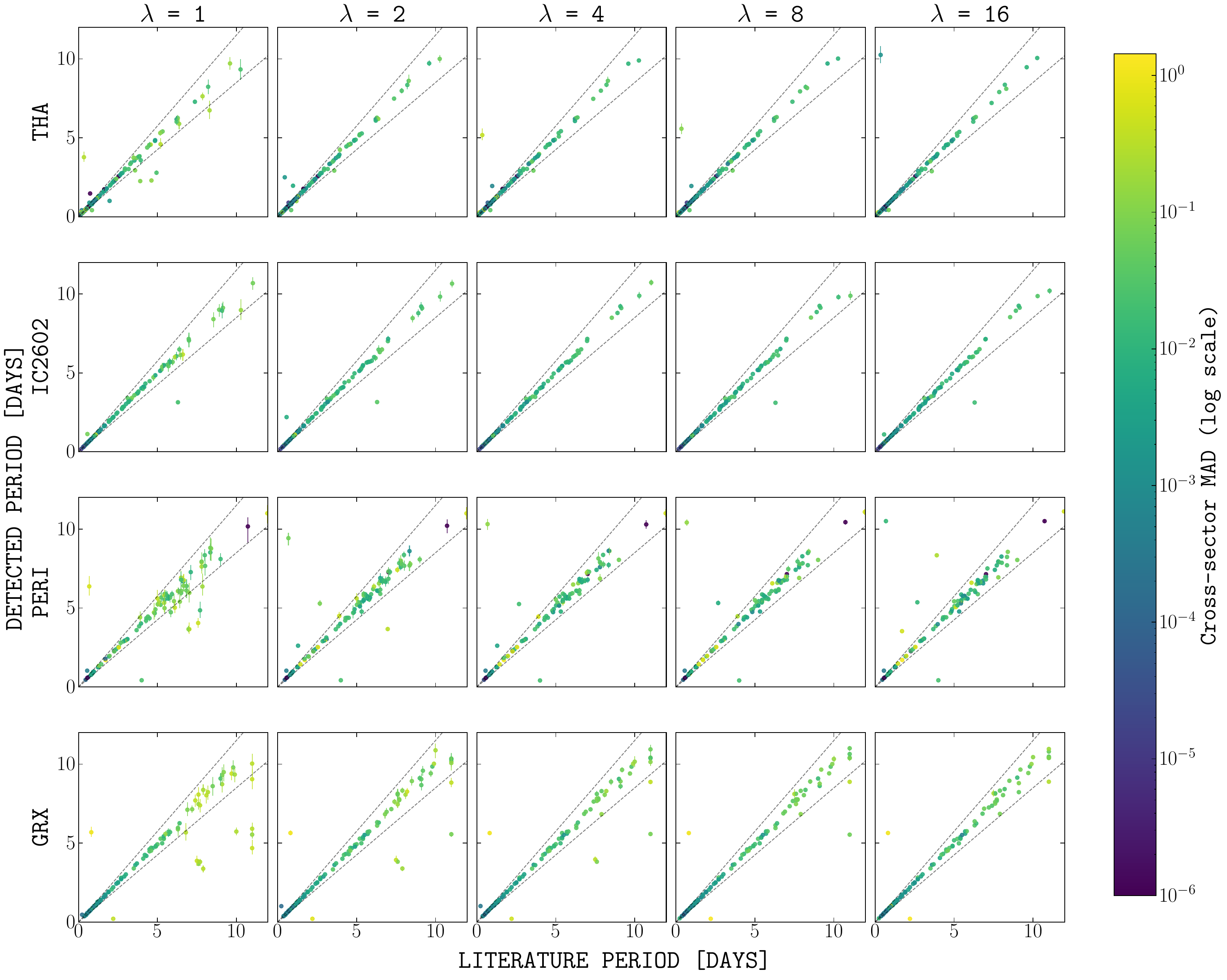} 
\caption{Comparison of literature rotation periods with the rotation periods detected for stars Tucana--Horologium Association, IC~2602, Pisces--Eridanus and Group~X. Dashed bars indicate the 15\% range around the literature periods. The color scale represents the cross-sector median absolute deviation in log scale.
.}\label{fig:multisector_consistency}
\end{figure*}

Figure~\ref{fig:multisector_consistency} compares the adopted GP periods with the literature values and colors each target by$D_{\mathrm{sector}}$. In addition to reducing several strongly discrepant period solutions, regularization generally decreases the smaller sector-to-sector offsets that are not apparent from the literature comparison alone. The cluster-level median of $D_{\mathrm{sector}}$ is lower than the unregularized value for every tested $\lambda>1$. Between $\lambda=1$ and $\lambda=16$, the decrease correspond to reductions of approximately 71\% for Tucana--Horologium Associtation, 51\% for IC2602, 62\% for Pisces--Eridanus, and 35\% for Group~X.

\tablewidth{1\textwidth}
\begin{deluxetable}{lrrrrrr}[!ht]
\tablecolumns{7}
\tablecaption{Median normalized cross-sector MAD of detected periods, computed as $\mathrm{MAD}(P)/\mathrm{median}(P)$, for each cluster and $\lambda$.\label{tab:median_normalized_mad}}
\tablehead{
\colhead{Cluster} &
\colhead{$\lambda=1$} &
\colhead{$\lambda=2$} &
\colhead{$\lambda=4$} &
\colhead{$\lambda=8$} &
\colhead{$\lambda=16$} 
}
\startdata
GRX    & 0.0114 & 0.0098 & 0.0084 & 0.0097 & 0.0074 \\
IC2602 & 0.0085 & 0.0058 & 0.0054 & 0.0050 & 0.0041 \\
PERI   & 0.0302 & 0.0171 & 0.00129 & 0.0145 & 0.00116 \\
THA    & 0.0024 & 0.0016 & 0.0008 & 0.0007 & 0.0007 
\enddata
\end{deluxetable}

The decrease is not strictly monotonic with increasing regularization strength because the cross-sector statistic is calculated only from sectors that satisfy the adopted rotation-probability threshold of 0.7. As $\lambda$ changes, individual sectors may enter or leave the accepted sample, altering both the number of sector-level measurements and the resulting median absolute fractional deviation. The statistic therefore reflects changes in both the inferred periods and the subset of sectors considered reliable. Nevertheless, the overall reduction relative to the unregularized case indicates that regularization generally improves the consistency of the recovered period across independent TESS sectors.

\section{Conclusion}
In this work, we have shown that the standard application of Gaussian Processes to high-cadence, high-precision TESS photometry can lead to physically misleading hyperparameter estimates, even when the model provides an excellent statistical fit to the data. In particular, we find that the marginal likelihood can be driven primarily by the covariance-complexity term rather than by meaningful improvements in the data-fit term. In this regime, the GP may converge to a period hyperparameter that describes a statistically preferred covariance structure but does not correspond to the physical stellar rotation period.

We introduced an explicit regularization factor, $\lambda$, that reweights the GP marginal likelihood and tested its effect on rotation-period inference for \totalsample stars in four young stellar populations: IC~2602, the Tucana--Horologium Association, Pisces--Eridanus, and Group~X. Because these stars have independently reported rotation periods, this benchmark isolates the question of GP hyperparameter interpretability while also testing automated period recovery in a curated sample. Relative to the standard GP likelihood, moderate regularization improves agreement with literature periods and reduces the occurrence of spurious harmonic or nonphysical solutions, particularly for stars with non-sinusoidal or evolving rotational modulation.

We also examined the consistency of the inferred periods across independent TESS sectors. This multi-sector analysis shows that regularization generally reduces the median absolute fractional deviation of the recovered periods across sectors, indicating that its effect is not limited to rare catastrophic failures. Instead, regularization also improves the reproducibility of smaller period offsets between independent light-curve segments.

These results demonstrate that GP hyperparameters should not be interpreted as physical stellar properties solely because the GP reproduces the observed light curve. Likelihood regularization provides a useful diagnostic for testing the stability and physical robustness of GP-inferred rotation periods. At the same time, the automated framework developed here recovers rotation periods with minimal human intervention and reaches broadly consistent results with manually vetted literature analyses for this benchmark sample. The present work should therefore be interpreted as a validation of regularized GP-based rotation recovery for curated or well-characterized samples, and a promising avenue for fully blind rotation-search pipeline for arbitrary field populations. Within this scope, regularized GP modeling provides a practical path toward more physically interpretable and scalable stellar-variability inference in large photometric surveys.

%Validation with \texttt{butterpy} simulations further confirmed that regularization effectively prevents the model from converging to spurious solutions or harmonics.

\begin{acknowledgments}
This work was supported by the Natural Sciences and Engineering Research Council of Canada (NSERC) (Program reference: 589653). This research has made use of data from the {\it Transiting Exoplanet Survey Satellite} (TESS) mission, obtained from the Mikulski Archive for Space Telescopes (MAST). All TESS data used in 1164 this paper can be found at: \dataset[10.17909/fwdt-2x66]{10.17909/fwdt-2x66}. We acknowledge the TESS team and the MAST staff for their efforts in collecting and providing these valuable datasets. We also thank the anonymous referee for constructive comments that improved this manuscript, and our colleagues and collaborators, such as Rocio Kiman for helpful discussions throughout this work. This research has benefited from the use of community-developed software and analysis tools. JF acknowledges support from NSF CAREER award number 2238468. JF and MP acknowledge support from NASA awards G06201 and G05120.  JMV acknowledges support from a Royal Society - Research Ireland University Research Fellowship (URF/1/221932, RF/ERE/221108). A.L. acknowledges support from the Fonds de recherche du Québec - Secteur Nature et technologies (FRQ-NT) under file \#349961

\end{acknowledgments}


\software{\texttt{celerite2} \citep{celerite2}, \texttt{Astropy} \citep{Astropy2022}, \texttt{Matplotlib} \citep{Hunter2007}, \texttt{NumPy} \citep{harris2020array}, \texttt{pandas} \citep{pandas2020}, \texttt{TESScut} \citep{Brasseur2019}, \texttt{unpopular} \citep{Hattori2021},
OpenAI. (2026). ChatGPT (Version 5.5) [Large language model]. https://chat.openai.com/chat }

\appendix
\section{Entropy and its connection to the GP marginal likelihood}\label{app:entropy}

\subsection{Entropy of a multivariate Gaussian and relation to GP complexity}\label{subsec:math_entropy}

In information theory, entropy quantifies the uncertainty associated with a probability distribution. For a continuous probability density $p(x)$, the Shannon entropy \citep{Shannon1948} is defined as:

\begin{equation}
    H[p(x)] = -\int p(x)\,\log p(x)\,dx .
\end{equation}

In Gaussian Process (GP) regression, the likelihood of the data is modeled as a multivariate normal distribution whose covariance matrix $\boldsymbol{\Sigma}$ is fully specified by the GP hyperparameters. The entropy of this distribution therefore provides a natural, kernel-independent measure of the complexity of the GP model. Let $x \sim \mathcal{N}(\boldsymbol{\mu}, \boldsymbol{\Sigma})$ be an $N$-dimensional multivariate Gaussian with probability density
\begin{equation}
    p(x) = (2\pi)^{-N/2} |\boldsymbol{\Sigma}|^{-1/2}
    \exp\!\left[-\frac{1}{2}(x-\boldsymbol{\mu})^\mathsf{T}
    \boldsymbol{\Sigma}^{-1}(x-\boldsymbol{\mu})\right].
\end{equation}

Substituting this expression into the entropy definition yields
\begin{equation*}
    H(x) = - \int_x p(x)\,\log{p(x)}\, dx,\text{ where } 
    p(x) = \mathcal{N}(x \mid \mu, \Sigma) = (2 \pi)^{-D/2}\, \mid \Sigma\mid^{-1/2} e^{-1/2 (x-\mu)^T \Sigma^{-1} (x-\mu)} dx
\end{equation*}
\\
\noindent In this equation, $\mu$ represents the mean, $\Sigma$ the covariance matrix, and $N$ is the dimensionality of the covariance matrix ($N \times N$).

\begin{equation*}
    H(x) = - \int_x \mathcal{N}(x \mid \mu, \Sigma) \cdot \left(-\dfrac{N}{2}\log{2 \pi} - \dfrac{1}{2} \log{\mid \Sigma\mid} -\dfrac{1}{2}(x-\mu)^T \Sigma^{-1} (x-\mu)\right) dx
\end{equation*}

\begin{equation*}
    H(x) = \dfrac{N}{2}\log{2 \pi}\int_x \mathcal{N}(x \mid \mu, \Sigma) dx  
    + \dfrac{1}{2} \log{\mid \Sigma\mid} \int_x \mathcal{N}(x \mid \mu, \Sigma) dx  +\dfrac{1}{2}\int_x \mathcal{N}(x \mid \mu, \Sigma) \left((x-\mu)^T \Sigma^{-1} (x-\mu)\right)dx
\end{equation*}
\\
\noindent By definitions, $\int_x \mathcal{N}(x \mid \mu, \Sigma) dx  = 1$, so we get:

\begin{equation*}
    H(x) = \dfrac{N}{2}\log{2 \pi} + \dfrac{1}{2} \log{\mid \Sigma\mid}  +\dfrac{1}{2}\int_x \mathcal{N}(x \mid \mu, \Sigma) \left((x-\mu)^T \Sigma^{-1} (x-\mu)\right)dx
\end{equation*}
\\
\noindent We can define the expected value $E[(x-\mu)^T \Sigma^{-1} (x-mu)] = \int_x (x-\mu)^T \Sigma^{-1} (x-mu) \, f((x-\mu)^T \Sigma^{-1} (x-\mu)) dx$, where $f(x) = \mathcal{N}(x\mid \mu, \Sigma)$

\begin{equation*}
    H(x) = \dfrac{N}{2}\log{2 \pi} + \dfrac{1}{2} \log{\mid \Sigma\mid}  +\dfrac{1}{2} E[(x-\mu)^T \Sigma^{-1} (x-\mu)]
\end{equation*}
\\
\noindent Considering that $(x -\mu) \Sigma^{-1} (x - \mu)$ is a scalar, we can rewrite our equation such that:

\begin{equation*}
    H(x) = \dfrac{N}{2}\log{2 \pi} + \dfrac{1}{2} \log{\mid \Sigma\mid}  +\dfrac{1}{2} E[tr\left((x-\mu)^T \Sigma^{-1} (x-\mu)\right)]
\end{equation*}
\\
\noindent We can leverage trace operator properties, such as  $tr(ABC) = tr(BCA)$, yielding:

\begin{equation*}
    H(x) = \dfrac{N}{2}\log{2 \pi} + \dfrac{1}{2} \log{\mid \Sigma\mid}  +\dfrac{1}{2} E[tr\left(\Sigma^{-1}(x-\mu)^T (x-\mu)\right)]
\end{equation*}
\begin{equation*}
    H(x) = \dfrac{N}{2}\log{2 \pi} + \dfrac{1}{2} \log{\mid \Sigma\mid}  +\dfrac{1}{2} tr(\Sigma^{-1})\, \left(E[(x-\mu)^T (x-\mu)]\right)
\end{equation*}
\\
\noindent
We know that the covariance is defined by $\Sigma = Cov((x- mu) , (x - \mu)) = E[(x-\mu)(x-\mu)^T] -  E[x - \mu)]E[(x - \mu)]^T$. However, by definition, $E[x - \mu)] = E[x - \mu)]^T = 0$, so we have that $\Sigma  = E[(x-\mu)(x-\mu)^T]$.

\begin{equation*}
    H(x) = \dfrac{N}{2}\log{2 \pi} + \dfrac{1}{2} \log{\mid \Sigma\mid}  +\dfrac{1}{2} tr(\Sigma^{-1}\,\Sigma)
\end{equation*}

\begin{equation*}
    H(x) = \dfrac{N}{2}\log{2 \pi} + \dfrac{1}{2} \log{\mid \Sigma\mid}  +\dfrac{1}{2} tr(\mathbb{I}_{N  \times N})
\end{equation*}

\begin{equation*}
    H(x) = \dfrac{N}{2}\log{2 \pi} + \dfrac{1}{2} \log{\mid \Sigma\mid}  +\dfrac{N}{2}
\end{equation*}

\begin{equation*}
    H(x) = \dfrac{N}{2}(1+\log{2 \pi}) + \dfrac{1}{2} \log{\mid \Sigma\mid} 
\end{equation*}

The entropy of the multivariate Gaussian therefore reduces to
\begin{equation}
    H(x)
    =
    \frac{N}{2}\log(2\pi e)
    +
    \frac{1}{2}\log|\boldsymbol{\Sigma}|
\end{equation}

Up to an additive constant independent of the GP hyperparameters, the entropy is therefore proportional to the log-determinant of the covariance matrix. Since $\log|\boldsymbol{\Sigma}|$ quantifies the effective volume of the covariance ellipsoid, it naturally acts as a measure of model complexity, as established by \cite{Rasmussen2006}

Importantly, the same log-determinant term appears in the Gaussian Process marginal log-likelihood:

\begin{equation}
    \log \hat{\mathcal{L}}
    =
    -\frac{1}{2}\log|\boldsymbol{\Sigma}|
    -
    \frac{1}{2}\mathbf{y}^\mathsf{T}
    \boldsymbol{\Sigma}^{-1}
    \mathbf{y}
    - \text{const}.
\end{equation}
This formal correspondence shows that the complexity penalty in the GP marginal likelihood is mathematically equivalent, up to additive constants, to minimizing the entropy of the Gaussian likelihood.

\subsection{Interpretation of entropy-based regularization}
In the standard Gaussian Process marginal likelihood, the log-determinant term enters with a negative sign, such that increasing $\log|\boldsymbol{\Sigma}|$ reduces the overall likelihood. As shown in Appendix~\ref{subsec:math_entropy}, this term is formally equivalent, to the entropy of the Gaussian likelihood. Consequently, the standard GP formulation implicitly favors covariance structures with lower entropy.

In our entropy-based interpretation, the regularization term enters with a positive sign when written explicitly as an entropy contribution, which may appear counterintuitive at first glance. However, when combined with the negative sign already present in the marginal likelihood, the net effect is to further penalize high-entropy covariance matrices. This effectively strengthens the preference for structured, low-entropy models.

Encouraging low entropy in this context corresponds to favoring covariance structures that encode coherent correlations over longer timescales, rather than highly flexible models dominated by short-scale variability. While high-entropy models are capable of fitting noise or isolated outliers, they tend to obscure physically meaningful signals and degrade extrapolative power. By contrast, low-entropy covariance matrices impose stronger correlations between data points, improving robustness to outliers and enhancing the interpretability of the inferred hyperparameters.

It is important to note that encouraging low-entropy  model reflects a controlled bias toward the simplest covariance structures within a flexible non-parametric framework. The regularization weight governs the strength of this bias, allowing one to prevent overfitting.

\section{How to implement this method in your favorite GP package}\label{sec:packages_m}
\subsection{\texttt{celerite, celerite2}}
For this section, we implement our regularization such that $M = \lambda$.
\begin{lstlisting}
import celerite2 
# instead of log_likelihood = gp.log_likelihood(flux) 
log_likelihood = M*gp._norm - 0.5 * gp._do_norm(flux - gp._mean_value) # M is the weight to optimize
\end{lstlisting}

\subsection{\texttt{george}}
\begin{lstlisting}
import george 
# sub-computations
mu = gp._call_mean(x)
r = np.ascontiguousarray( gp._check_dimensions(y) - mu, dtype=np.float64)
# instead of log_likelihood = gp.log_likelihood(flux) 
log_likelihood = M * self._const - 0.5 * self.solver.dot_solve(r) # M is the weight to optimize
\end{lstlisting}

\subsection{\texttt{tinygp}}

\begin{lstlisting}
import tinygp
import jax.numpy as jnp
#sub-computations
alpha = gp._get_alpha(y)
#instead of _, log_likelihood, _ = gp.condition(y, x)
log_likelihood = -0.5 * jnp.sum(jnp.square(alpha)) - M*self.solver.normalization() # M is the weight to optimize
\end{lstlisting}

\bibliography{moranta26_bib}
\bibliographystyle{aasjournalv7}



\end{document}